\pdfoutput=1
\documentclass[%
 reprint,
superscriptaddress,
 amsmath,amssymb,
 aps,
prb,
floatfix,
]{revtex4-1}
\usepackage[T1]{fontenc}
\usepackage{algorithm2e}
\usepackage{graphicx}
\usepackage{dcolumn}
\usepackage{bm}
\usepackage{hyperref}
\usepackage{xcolor}
\usepackage{lipsum}

\newcommand{\diff}{\mathop{}\!\mathrm{d}}
\renewcommand{\vec}[1]{\mathbf{#1}}

\DeclareMathOperator{\Tr}{Tr}

\begin{document}

\preprint{APS/123-QED}

\title{Phase field crystal model for heterostructures}

\author{Petri Hirvonen}
\email{petenez@gmail.com}
\affiliation{QTF Centre of Excellence, Department of Applied Physics, Aalto University School of Science, P.O. Box 11000, FIN-00076, Aalto, Espoo, Finland}

\author{Vili Heinonen}
\affiliation{Department of Mathematics, Massachusetts Institute of Technology, 77 Massachusetts Avenue, Cambridge,~MA~02139-4307, USA}

\author{Haikuan Dong}
\affiliation{School of Mathematics and Physics, Bohai University, Jinzhou 121000, China}

\author{Zheyong Fan}
\affiliation{QTF Centre of Excellence, Department of Applied Physics, Aalto University School of Science, P.O. Box 11000, FIN-00076, Aalto, Espoo, Finland}

\author{Ken R. Elder}
\affiliation{Department of Physics, Oakland University, Rochester, MI 48309, USA}

\author{Tapio Ala-Nissila}
\affiliation{QTF Centre of Excellence, Department of Applied Physics, Aalto University School of Science, P.O. Box 11000, FIN-00076, Aalto, Espoo, Finland}
\affiliation{Interdisciplinary Centre for Mathematical Modelling and Department of Mathematical Sciences, Loughborough University, Loughborough, Leicestershire LE11 3TU, UK}

\date{\today}

\begin{abstract}
Atomically thin 2-dimensional heterostructures are a promising, novel class of materials with groundbreaking properties. 
The possiblity of choosing the many constituent components and their proportions allows optimizing these materials to specific requirements. 
The wide adaptability comes with a cost of large parameter space making it hard to experimentally test all the possibilities. 
Instead, efficient computational modelling is needed. 
However, large range of relevant time and length scales related to physics of polycrystalline materials poses a challenge for computational studies. 
To this end, we present an efficient and flexible phase-field crystal model to describe the atomic configurations of multiple atomic species and phases coexisting in the same physical domain. We extensively benchmark the model for two-dimensional binary systems in terms of their elastic properties and phase boundary configurations and their energetics. As a concrete example, we demonstrate modelling lateral heterostructures of graphene and hexagonal boron nitride. We consider both idealized bicrystals and large-scale systems with random phase distributions. We find consistent relative elastic moduli and lattice constants, as well as realistic continuous interfaces and faceted crystal shapes. Zigzag-oriented interfaces are observed to display the lowest formation energy.
\end{abstract}

\maketitle


\section{Introduction}

Most scientifically and technologically important materials are composed of multiple atomic species and phases with different chemical compositions,  lattice structures and elastic properties. Some everyday examples include wood, rock, metallic alloys, and concrete. Such three-dimensional (3D) materials have been used for thousands of years and efforts towards their development continue in the age of nanophysics. Some modern examples include e.g. fiber-reinforced polymers and semiconductor heterostructures. 
The past decade has seen the emergence of a completely new type of materials, the atomically thin two-dimensional (2D) materials. The extraordinary properties of single component 2D materials \cite{ref-properties-1, ref-properties-2, ref-properties-3, ref-properties-4, ref-properties-5, ref-properties-6, ref-properties-7, ref-properties-8, ref-properties-9, ref-our-pccp18} can be widely enhanced and adjusted by considering their heterostructures that can either be stacked to form vertical multilayer heterostructures \cite{ref-heterostructures-vertical-1, ref-heterostructures-vertical-2, ref-heterostructures-vertical-3}, or they can be grown within a single material sheet into a lateral heterostructure \cite{ref-heterostructures-lateral-1,ref-heterostructures-lateral-2,ref-heterostructures-lateral-3,ref-heterostructures-lateral-4}.

The properties of pure or single phase crystalline materials are determined by the complex networks of microscopic defects and grains. In contrast, for many multiphase composite materials, macroscopic continuum models may provide sufficiently accurate predictions of many of their properties. This suggests that the role of their microscopic structure is less important. Nevertheless, for semiconductor heterostructures, as well as for vertical and lateral 2D heterostructures, the atomic-level structure of their phase interfaces plays a major role as said structures are miniaturized to the nanoscale where interfacial effects are important.

Predicting the atomic-level structure between two or more orientationally, structurally and elastically mismatched phases is particularly difficult. The number of possible atomic configurations is essentially endless and conventional atomistic modelling techniques cannot simultaneously reach all the length and time scales relevant to their formation. The more recently developed phase field crystal (PFC) method allows examination of long, diffusive time scales corresponding to the slow evolution of microstructures, and offers atomic-level spatial resolution up to mesoscopic length scales \cite{ref-pfc-2002}. PFC models describe crystalline matter in terms of smooth, classical density fields $n_i$ of the different atomic species. The essential thermodynamic quantity is the free energy $F \left[ n_i \right]$ that is minimized by a periodic $n_i$. 
PFC models have been extensively applied to study various complex systems and processes such as grain boundaries, vacancy diffusion, coarsening of polycrystals, heteroepitaxial growth, yield strength and fracture \cite{ref-pfc-2002, ref-pfc-2004, ref-pfc-review-2012, ref-pf-pfc-review-2019}. In particular, multicomponent PFC models that explicitly incorporate multiple density fields $n_i$ coupled together, have been developed and applied to study crystal structures composed of multiple atomic species \cite{ref-pfc-multicomponent-1, ref-pfc-multicomponent-2, ref-pfc-multicomponent-3, ref-pfc-multicomponent-4, ref-pfc-h-bn,carter17}. 

In this work we introduce \emph{spatially smoothed atomic density fields} coupled to the atomic density fields $n_i$ that enables well-controlled phase separation and, therefore, facilitates modelling heterostructures and composite materials. Smoothed densities have been employed in PFC modelling recently for introducing a vapor phase \cite{ref-pfc-vapor} and for  controlling liquid/solid interface energies \cite{ref-pfc-controlled-interface-energies}. Here we apply this modelling approach to 2D heterostructures composed of multiple elements. We carry out a systematic investigation by varying model parameters one by one to determine their influence on the general behavior of the model. More specifically, we introduce mismatch in both the elastic moduli and the lattice constants between the two materials, as well as experiment with different couplings between the two density fields. Finally, we assess the model's suitability to study graphene--hexagonal boron nitride 2D heterostructures.

The rest of the paper is organized as follows: Section \ref{sec-model} lays out and discusses the heterostructure model and gives some practical details of our calculations. Section \ref{sec-binary} presents our investigation of the general properties of the model using binary heterostructures. Section \ref{sec-trinary} assesses the model's suitability to modelling graphene--hexagonal boron nitride lateral heterostructures. Finally, Sec. \ref{sec-conclusions} summarizes the work.

\section{Heterostructure model}

\label{sec-model}

Phase field crystal (PFC) models are a family of continuum methods for multiscale modelling of polycrystalline materials and their complex microstructures. PFC models allow simultaneous access to both atomic and mesoscopic length scales, as well as to long, diffusive time scales.
Formation and evolution of microstructures take place in this time regime. Conventional PFC models use a smooth, periodic density field $n$ to describe crystalline systems. The length scale and lattice symmetries, as well as the elastic properties of the model can be matched with the target material. These properties are determined by a free energy functional $F\left[ n \left( \vec{r} \right) \right]$ governing the energetics of the system. In the solid phase $F$ is minimized by a periodic $n$ whose symmetries depend on the formulation of $F$ and average density $\bar{n}$ \cite{ref-pfc-2002, ref-pf-book}. A PFC model can incorporate multiple density fields coupled together to allow the study of more complex structures. Such models have been applied to study multicomponent materials such as 2D hexagonal boron nitride (h-BN) \cite{ref-pfc-h-bn}.

In conventional multicomponent PFC models the periodically oscillating densities representing the solid phase of each component overlap and form interlocking, mixed lattices. To study heterostructures with well-controlled phase separation, we propose the following dimensionless free energy functional:

\begin{equation}
    \label{eq-model}
    \begin{split}
        F = \int \diff \vec{r} \left( \sum_{i = 1}^N \left( \frac{\alpha_i}{2} n_i^2 + \frac{\beta_i}{2} n_i \left( \nu_i^2 + \nabla^2 \right)^2 n_i \right. \right.
        \\
        \left. + \frac{\gamma_i}{3} n_i^3 + \frac{\delta_i}{4} n_i^4 \right)
        \\
        + \sum_{i = 1}^{N - 1} \sum_{j = i + 1}^N \left( \alpha_{ij} n_i n_j \vphantom{\frac{\gamma_{ij}}{2}} + \beta_{ij} n_i \left( \nu_{ij}^2 + \nabla^2 \right)^2 n_j \right.
        \\
        \left. \left. + \frac{\gamma_{ij}}{2} \left( n_i^2 n_j + n_i n_j^2 \right) + \epsilon_{ij} \eta_i \eta_j \vphantom{\frac{\gamma_{ij}}{2}} \right) \vphantom{\sum_{i = 1}^N} \right).
    \end{split}
\end{equation}

\noindent Here, the first sum contains the ideal contributions of the $N$ density fields and the second, nested sum the contributions of the interactions between them. In the first sum, the quadratic and quartic terms comprise a double-well potential, the cubic term acts similarly to a chemical potential and the gradient term gives rise to periodic solutions and elastic behavior. We refer the reader to Refs. \cite{ref-pfc-2002, ref-pf-book, ref-akusti-thesis, ref-vili-thesis} for a more in-depth discussion of PFC formulation. In the second sum, the quadratic and cubic terms are local couplings between the different density fields, whereas the rest are nonlocal terms. The model parameters and their roles are  summarized in Table \ref{tab-parameters}.

The term $\int \epsilon_{ij} \eta_i \eta_j \diff \vec{r}$ in Eq. (\ref{eq-model}) is the essential coupling responsible for controlled phase separation. This term can effectively drive $\bar{n}_i$ and $\bar{n}_j$ apart in the same physical domain such that one corresponds to the disordered and the other to a crystalline phase in the phase diagram. 
The fields $\eta_i$ are spatially smoothed $n_i$ where the atomic-level structures have been filtered out defined as
$\eta_i = G \ast n_i$. Here the asterisk denotes a convolution and $G$ is a Gaussian smoothing kernel with the Fourier transform $\hat{G} \left( \vec{k} \right) = e^{- \left| \vec{k} \right|^2/\left( 2 \sigma^2 \right)}$.
In the present work we found that $\sigma = 0.2$ corresponding to a length scale of approximately five lattice constants with $\nu_i = 1$
sufficiently smooths out the atomic-level structure. To enable atomistically sharp interfaces and, moreover, to keep the number of parameters to be tuned to a minimum, we did not consider values $\sigma < 0.2$. 
The influence of this coupling term is demonstrated in Sec. \ref{sec-epsilon} and its ability to drive phase separation is shown analytically in Appendix \ref{sec-appendix-a}.

In order to find the $n_i$ that minimize $F$, we used density conserving gradient descent given by 
\begin{equation}
    \label{eq-dynamics}
    \begin{split}
        \frac{\partial n_i}{\partial t} = &\nabla^2 \frac{\delta F}{\delta n_i}
        \\
        + \nabla^2 &\left( \vphantom{\sum_{\substack{j = 1 \\ j \neq i}}^N} \alpha_{i} n_i + \beta_i \left( \nu_i^2 + \nabla^2 \right)^2 n_i + \gamma_i n_i^2 + \delta_i n_i^3 \right.
        \\
        &+ \sum_{\substack{j = 1 \\ j \neq i}}^N \left( \vphantom{\frac{\gamma_{ij}}{2}} \alpha_{ij} n_j + \beta_{ij} \left( \nu_{ij}^2 + \nabla^2 \right)^2 n_j \right.
        \\
        &\left. \left. + \frac{\gamma_{ij}}{2} \left( 2 n_i n_j + n_j^2 \right) + \epsilon_{ij} G \ast \eta_j \right) \vphantom{\sum_{\substack{j = 1 \\ j \neq i}}^N} \right)
    \end{split}
\end{equation}


\noindent where $\alpha_{ij} = \alpha_{ji}$ and similarly for the other parameters. The density conservation constraint is essential for stabilizing the heterostructures. Note that while $t$ is called sometimes \emph{time}, it is a relaxation parameter that is not related to real dynamics in this work.
We solved Eq.~\eqref{eq-dynamics} numerically using a semi-implicit spectral method described in Ref.~\citenum{ref-pf-book}. This method allows computing the gradients and convolutions present in Eqs. (\ref{eq-model}) and (\ref{eq-dynamics}) accurately and efficiently by using fast Fourier transform routines. 
Note that according to the convolution theorem, convolutions can be expressed as $f \ast g = \mathcal{F}^{-1} \left\lbrace \hat{f} \hat{g} \right\rbrace$, where $\mathcal{F}^{-1}$ and the carets indicate inverse and forward Fourier transforms, respectively. The following upper bounds for spatial and temporal discretizations were used for all the calculations: $\Delta x = 0.75$, $\Delta y = 0.75$, and $\Delta t = 0.25$.

Finally, we also used a model system size optimization algorithm \cite{ref-our-prb16} to eliminate strain in our bicrystalline model systems of heterostructures. We did not apply the method to polycrystalline systems, since we did not attempt to extract equilibrium densities from them or to analyze them quantitatively here.

\begin{table}
\centering
\caption{Summary of model parameters. They are listed below and their significance is explained. While not explicitly written in Eqs. (\ref{eq-model}) or (\ref{eq-dynamics}), the average densities $\bar{n}_i$ are important in controlling the relative stability of different phases.}
\label{tab-parameters}
\begin{tabular}{ c l }
\hline
\hline
 $N$ & number of density fields in the model \\
 $\bar{n}_i$ & average density; controls the relative stability of \\
  & different phases \\
 $\alpha_i$ & temperature-related parameter; controls the \\
  & diffuseness and facetedness of structures \\
 $\beta_i$ & controls the elastic moduli \\
 $\gamma_i$ & acts similarly to a chemical potential \\
 $\delta_i$ & usually set to unity in PFC models \\
 $\nu_i$ & wavenumber that sets the length scale \\
 $\alpha_{ij}$ & controls the alignment of two lattices at their mutual \\
  & interface but also introduces so-called "weak \\
  & oscillations" (see Sec. \ref{sec-alpha}) \\
 $\beta_{ij}$ & needed for h-BN \cite{ref-pfc-h-bn} \\
 $\gamma_{ij}$ & needed for h-BN \cite{ref-pfc-h-bn} \\
 $\epsilon_{ij}$ & couples smoothed densities; controls phase separation \\
$\nu_{ij}$ & needed for h-BN \cite{ref-pfc-h-bn} \\
 $\sigma$ & spectral spread of the Gaussian smoothing kernel $G$ \\
\hline
\hline
\end{tabular}
\end{table}

\section{Binary heterostructures}

\label{sec-binary}

We begin by demonstrating some general properties of the present model for simple binary heterostructures with $N = 2$ and denote the two density fields by $n_1$ and $n_2$.
We vary certain model parameters to investigate their influence and will refer to the periodic or ``crystalline regions'' in $n_i$ by $n_i^\textrm{(c)}$ and, similarly, to the disordered regions by $n_i^\textrm{(d)}$. The ``crystalline phase $i$'' encompasses regions where $n_i^\textrm{(c)}$ and $n_j^\textrm{(d)}$ coincide. Similarly, the ``mixed phase'' (disordered phase) spans the regions where $n_i^\textrm{(c)}$ and $n_j^\textrm{(c)}$ ($n_i^\textrm{(d)}$ and $n_j^\textrm{(d)}$) coincide.

\subsection{Influence of smoothed coupling}

\label{sec-epsilon}

The crucial parameter here is $\epsilon_{12}$ in the coupling term for the smoothed density fields. With $\epsilon_{12} > 0$, $n_1^\textrm{(c)}$ and $n_2^\textrm{(c)}$ repel each other, whereas with $\epsilon_{12} < 0$, $n_1^\textrm{(c)}$ and $n_2^\textrm{(c)}$ attract each other; see Appendix \ref{sec-appendix-a} for an analytical treatise. For the other parameters, we chose $\left( \alpha_i, \beta_i, \nu_i, \gamma_i, \delta_i, \alpha_{12}, \beta_{12}, \gamma_{12} \right) = \left( -0.3, 1.0, 1.0, 0.0, 1.0, 0.0, 0.0, 0.0 \right)$, for simplicity. This choice of model parameters is used throughout Sec. \ref{sec-binary} unless stated otherwise. Note that this choice of model parameters is symmetric, \textit{i.e.}, $F \left( n_1, n_2 \right) = F \left( n_2, n_1 \right)$.

\begin{figure*}
    \centering
    \includegraphics[width=\textwidth]{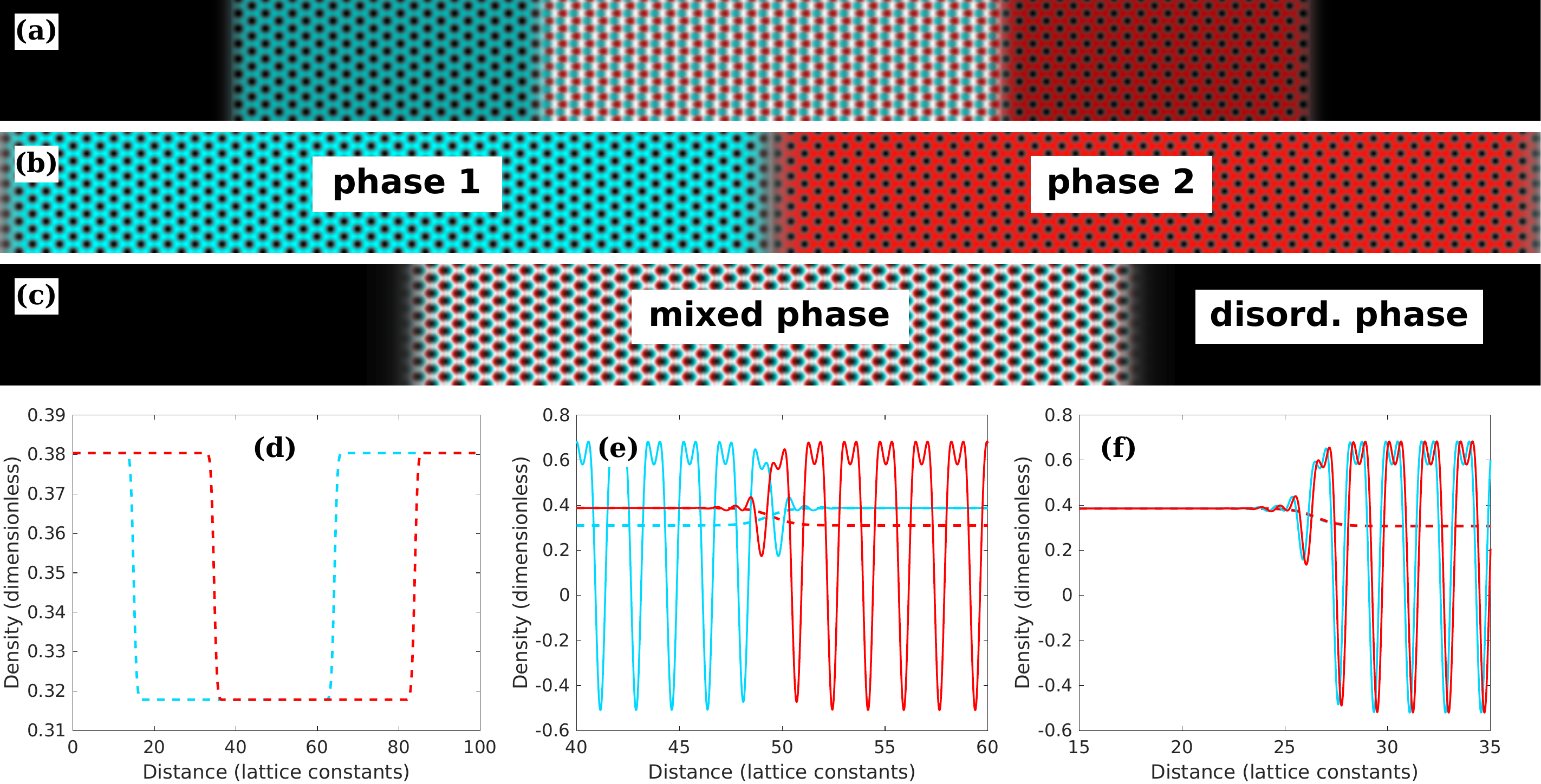}
    \caption{Influence of the coupling parameter $\epsilon_{12}$ on the heterostructures. (a) The initial state with $n_1^\textrm{(c)}$ and $n_2^\textrm{(c)}$ in partial overlap. (b) The relaxed heterostructure for the repulsive case where $\epsilon_{12} = 0.2$. Here, the crystalline phases 1 and 2 are shown in cyan and red, respectively. (c) The relaxed coexistence between a mixed and a disordered phase for the attractive case where $\epsilon_{12} = -0.2$. Here, the mixed phase appears white due to the coindicental alignment of the structures in $n_1^\textrm{(c)}$ and $n_2^\textrm{(c)}$, and the disordered phase appears black. (d) Profiles of the smoothed densities $\eta_1$ (cyan) and $\eta_2$ (red) along the periodic edge in the horizontal direction of the initial state from panel (a). (e) Profiles of the densities (solid lines) $n_1$ (cyan) and $n_2$ (red) and of the smoothed densities (dashed lines) $\eta_1$ (cyan) and $\eta_2$ (red) along the periodic edge in the horizontal direction of the relaxed heterostructure from panel (b). (f) Same profiles for panel (c).}
    \label{fig-epsilon-bi}
\end{figure*}

We used simple model systems where we initialized both $n_1$ and $n_2$ roughly 50\% crystalline and 50\% disordered with average densities $\bar{n}_i^\textrm{(c)} \approx 0.32$ and $\bar{n}_i^\textrm{(d)} \approx 0.38$. The initial structure for $n_i^\textrm{(c)}$ was obtained using an inverted hexagonal one-mode approximation \cite{ref-one-mode-approximations}. We also arranged $n_1^\textrm{(c)}$ and $n_2^\textrm{(c)}$ in partial overlap to force some changes in them during relaxation. Figure \ref{fig-epsilon-bi} illustrates the relaxed heterostructures obtained with $\epsilon_{12} = \pm 0.2$. Panels (a) - (c) visualize the systems and panels (d) - (f) plot corresponding density profiles. Panel (a) shows the initial state with $n_1^\textrm{(c)}$ and $n_2^\textrm{(c)}$ in partial overlap. Panel (b) gives the repulsive case with $\epsilon_{12} = 0.2$ where $n_1^\textrm{(c)}$ and $n_2^\textrm{(c)}$ have pushed themselves apart to eliminate the mixed phase. Panel (c) depicts the attractive case with $\epsilon_{12} = -0.2$ where $n_1^\textrm{(c)}$ and $n_2^\textrm{(c)}$ have come to a full overlap forming a coexistence between a mixed and a disordered phase. Recall that $n_1$ and $n_2$ are coupled here only via $\eta_1$ and $\eta_2$ whereby the two atomic lattices do not interact. Consequently, the lattices can end up arbitrarily aligned such as here; see panel (c).

\begin{figure*}
    \centering
    \includegraphics[width=0.75\textwidth]{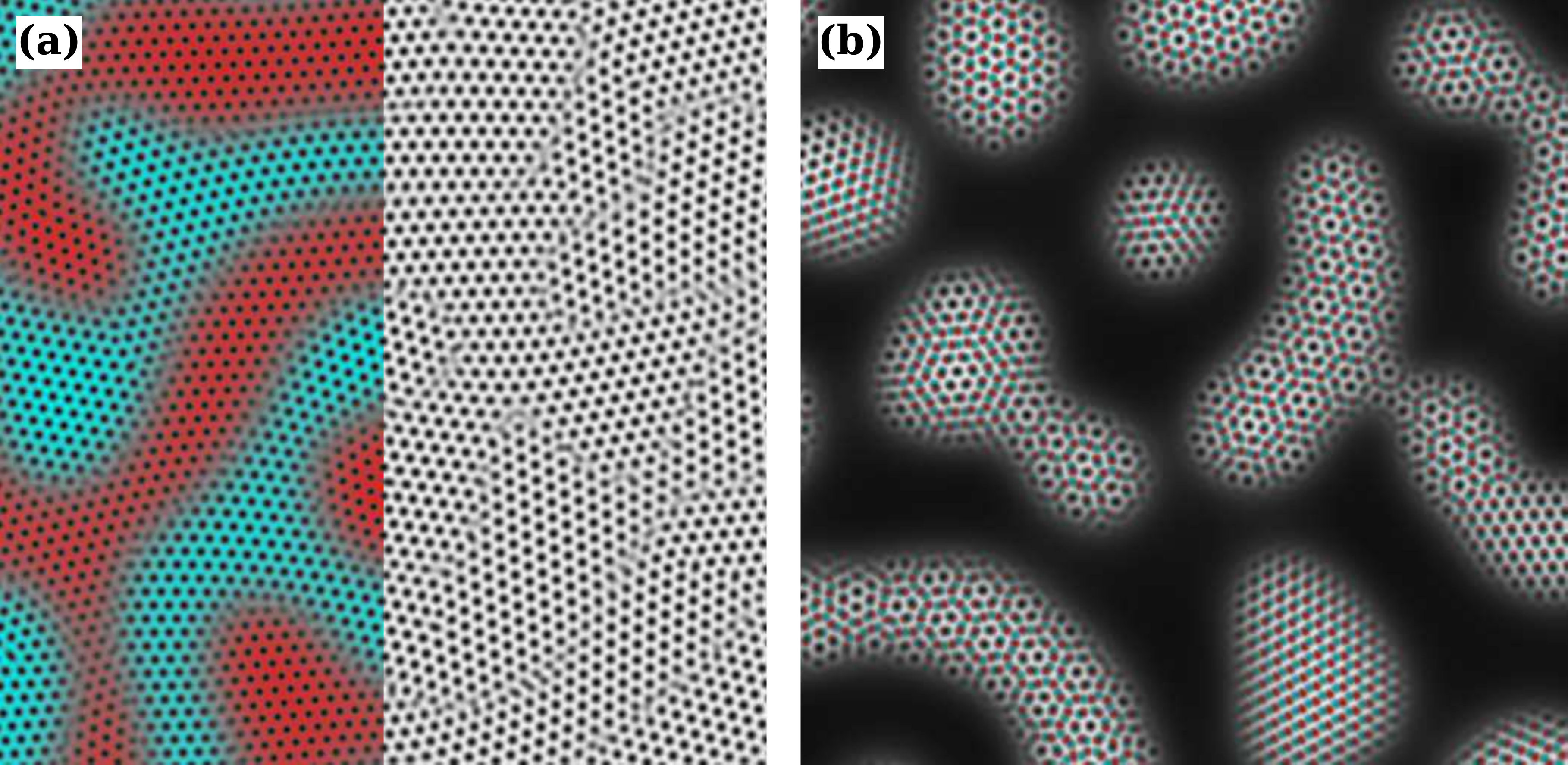}
    \caption{Influence of $\epsilon_{12}$ on polycrystalline heterostructures. (a) A blow-up of a larger system for the repulsive case where $\epsilon_{12} = 1.0$ after 7 500 time units. The left-hand side of the panel reveals the distribution of the two phases and the right-hand side represents the heterostructure by $m = n_1 + n_2$ for a clearer illustration of the atomic-level structure. (b) A blow-up of a larger system for the attractive case where $\epsilon_{12} = -1.0$ after 7 500 time units. Moir\'e overlap patterns between the two lattices are clearly visible.}
    \label{fig-epsilon-poly}
\end{figure*}

Next we considered polycrystalline heterostructures. The density fields $n_1$ and $n_2$ were initialized with white noise with $\bar{n}_i = 0.35$. With $\epsilon_{12} = 0.2$, a mixed phase emerges first, followed by delayed decomposition into the two separate crystalline phases. We, therefore, used $\epsilon_{12} = \pm 1.0$ to drive $n_1^\textrm{(c)}$ and $n_2^\textrm{(c)}$ apart ($+$) or together ($-$). Note, however, that if the coupling strength is increased significantly more, stripe phases \cite{ref-pfc-2002} may replace the crystalline ones as the most stable phase. Figure \ref{fig-epsilon-poly} demonstrates both the repulsive and attractive cases after 7 500 time units of relaxation. In the repulsive case shown in panel (a), $n_1^\textrm{(c)}$ and $n_2^\textrm{(c)}$ are well separated and there is no mixed or disordered phase. The interfaces between the two phases appear fuzzy and disordered as expected due to no interaction between the two underlying lattices. The attractive case is shown in panel (b) where $n_1^\textrm{(c)}$ and $n_2^\textrm{(c)}$ are in full overlap, yielding a patched coexistence between a mixed and a disordered phase. The arbitrary misorientations and translations between the lattices in $n_1^\textrm{(c)}$ and $n_2^\textrm{(c)}$ result in a multitude of Moir\'e patterns.

\subsection{Influence of \texorpdfstring{$\alpha_{ij}$}{aij}}

\label{sec-alpha}

Next, we varied the quadratic coupling parameter $\alpha_{12}$ to study its influence on the interfaces between the two crystalline phases and on the heterostructures as a whole. Here, we fix $\epsilon_{12} = 1.0$. We chose $\alpha_{12} < 0$ to achieve commensurate alignment of the two crystalline lattices at their interface to ensure the continuity of the underlying honeycomb lattice there. A side effect of this coupling is that it causes $n_i^\textrm{(c)}$ to induce oscillations in $n_j^\textrm{(d)}$. The amplitude of such weak oscillations in $n_j^\textrm{(d)}$ should be constrained to keep the two crystalline phases from mixing together. Indeed, the weak oscillations can be viewed as slight intermixing of the different atomic species. Intermixing is common in metallic alloys and in doped semiconductors and has been observed in lateral heterostructures of graphene and hexagonal boron nitride as well \cite{ref-intermixing}.

\begin{figure}
    \centering
    \includegraphics[width=0.4\textwidth]{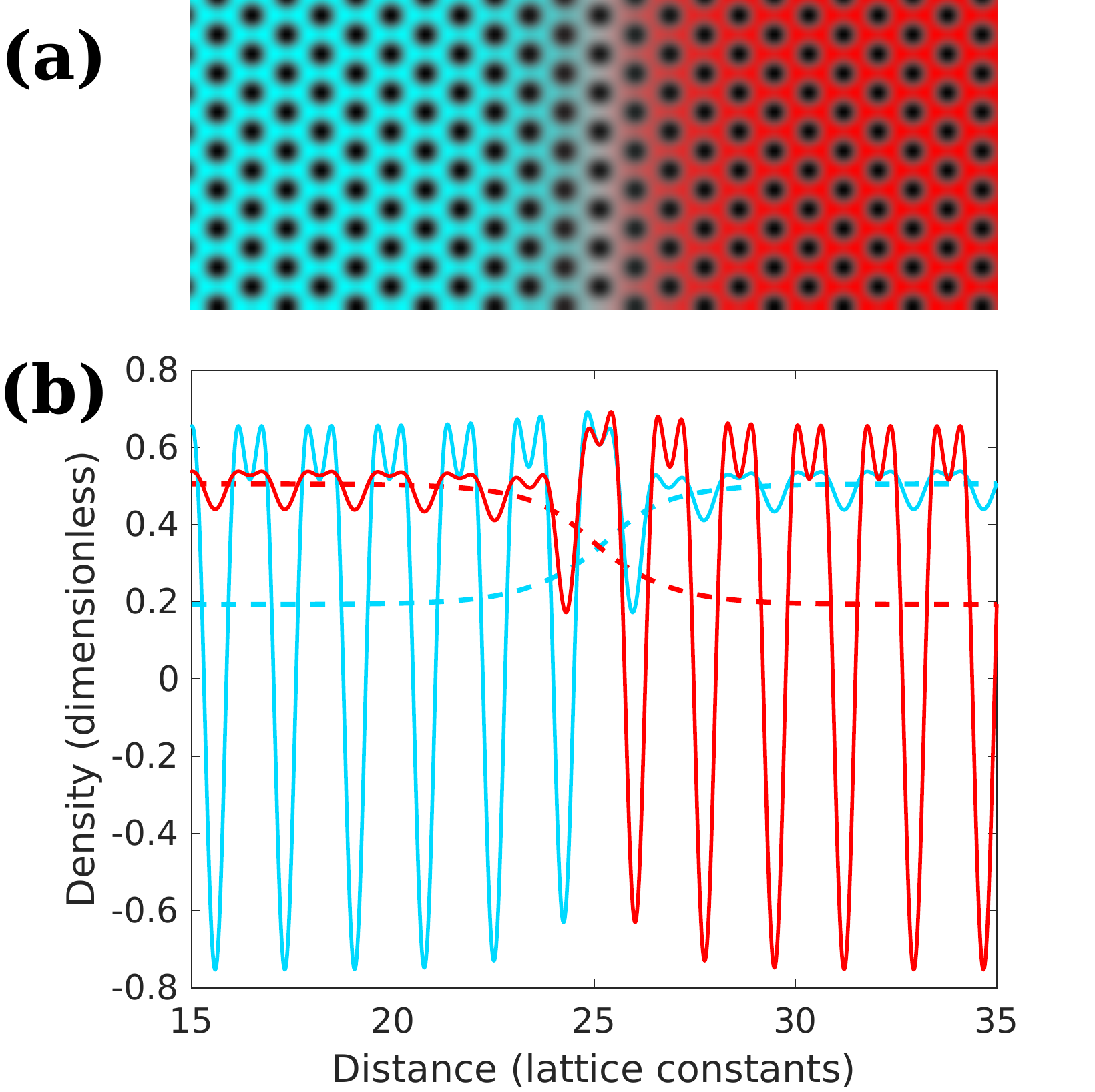}
    \caption{Influence of $\alpha_{12}$ and the corresponding coupling on a bicrystalline heterostructure. (a) A blow-up of a relaxed heterostructure with $\alpha_{12} = -0.03$. (b) Profiles of the densities (solid lines) $n_1$ (cyan) and $n_2$ (red) and of the smoothed densities (dashed lines) $\eta_1$ (cyan) and $\eta_2$ (red) along the periodic horizontal edge of the relaxed heterostructure from panel (a).}
    \label{fig-alpha-bi}
\end{figure}

While constrained by the amplitude of the weak oscillations induced, the magnitude of $\alpha_{12}$ should be maximized to ensure continuity even for lattice-mismatched or misoriented interfaces. We optimized $\alpha_{12}$ using bicrystalline heterostructures. We observed that the heterostructures are rendered unstable when $\alpha_{12} = -0.1$, but with $\alpha_{12} = -0.03$ they retain their stability while the amplitude of the weak oscillations remains negligible. Figure \ref{fig-alpha-bi} illustrates the interface in a relaxed bicrystalline heterostructure with $\alpha_{12} = -0.03$, $\bar{n}_i^\textrm{(c)} = 0.12$ and $\bar{n}_i^\textrm{(d)} = 0.58$. It is clear both from the visualization of the heterostructure as well as from the density profiles below it that the honeycomb lattice is highly continuous from one phase to the other. Furthermore, the interface is atomically sharp with an approximate width of two lattice constants.

\begin{figure*}
    \centering
    \includegraphics[width=0.75\textwidth]{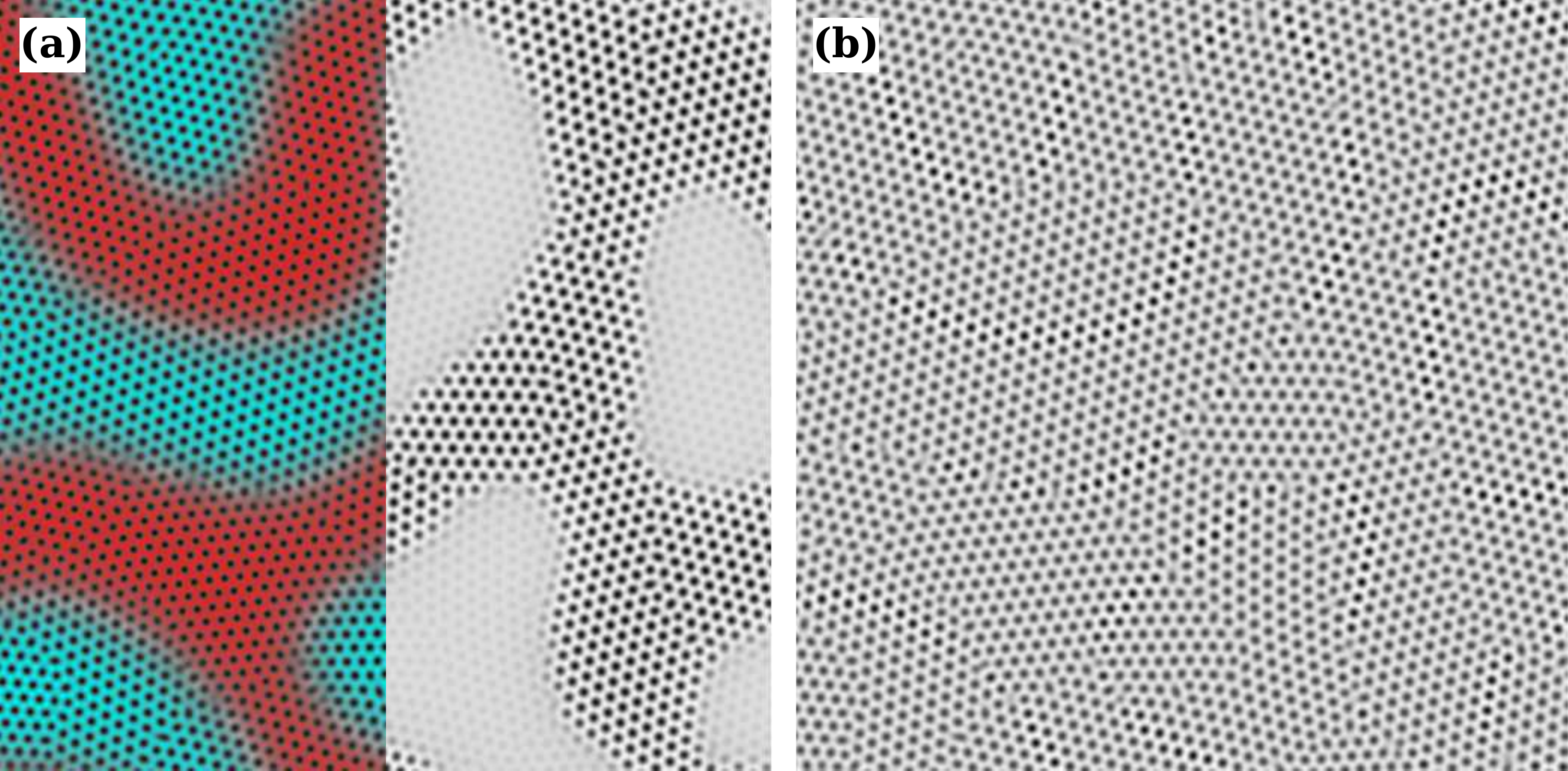}
    \caption{Influence of $\alpha_{12}$ on a polycrystalline heterostructure. (a) A blow-up of a larger system after a relaxation of 25 000 time units. The left-hand side of the panel reveals the distribution of the two phases and the right-hand side demonstrates $n_2$ with weak oscillations in $n_2^\textrm{(d)}$. (b) The heterostructure from panel (a) represented by $m = n_1 + n_2$ for a clearer illustration of the atomic-level structure.}
    \label{fig-alpha-poly}
\end{figure*}

We further demonstrated the influence of $\alpha_{12}$ for polycrystalline heterostructures. We initialized $n_1$ and $n_2$ with white noise where $\bar{n}_i = 0.35$. Figure \ref{fig-alpha-poly} demonstrates a relaxed heterostructure. On a larger scale, the system resembles that shown in Fig. \ref{fig-epsilon-poly} (a), but here the interfaces between the two crystalline phases are better ordered and more continuous. The width of the interfaces appears small for all misorientations. Note also the weak oscillations in $n_2^\textrm{(d)}$ visible in panel (a).

\subsection{Influence of \texorpdfstring{$\beta_i$}{bi}}

\label{sec-beta}

The crystalline phase $i$ is present where $n_i^\textrm{(c)}$ and $n_j^\textrm{(d)}$ coincide. The elastic properties of said phase should be dictated by $n_i$, but $n_j$ can have a minor contribution as well. We demonstrate here to what extent the elastic properties of the two crystalline phases can be controlled via $\beta_1$ and $\beta_2$, the coefficients of the gradient terms in $F$ responsible for elastic contribution. The ability to control the elastic stiffness of both phases separately is essential when modelling realistic heterostructures. Note that we show in Appendix \ref{sec-appendix-b} that the smoothed densities $\eta_i$ have a negligible elastic contribution.

The contribution from a uniform elastic deformation to the free energy density is given by

\begin{equation}
    \label{eq-elastic}
    f_\textrm{e} = \frac{C_{11}}{2} \left( \varepsilon_x^2 + \varepsilon_y^2 \right) + C_{12} \varepsilon_x \varepsilon_y,
\end{equation}

\noindent where $\varepsilon_x$ and $\varepsilon_y$ are the $x$ and the $y$ components of strain, and $C_{11} = C_{22}$ and $C_{12} = C_{21}$ are the stiffness coefficients. Furthermore, the bulk, shear and 2D Young's moduli, as well as Poisson's ratio are given by

\begin{equation}
    B = \frac{C_{11} + C_{12}}{2},
\end{equation}

\begin{equation}
    \mu = \frac{C_{11} - C_{12}}{2},
\end{equation}

\begin{equation}
    \label{eq-youngs-modulus}
    Y_\textrm{2D} = \frac{4 B \mu}{B + \mu},
\end{equation}

\noindent and

\begin{equation}
    \nu = \frac{B - \mu}{B + \mu},
\end{equation}

\noindent respectively \cite{ref-chaikin-lubensky}.

We determined the elastic coefficients of the two crystalline phases separately by straining single-crystals of either phase in the small deformation limit. More precisely, we varied $-0.002 \leq \varepsilon_x \leq 0.002$ and $-0.002 \leq \varepsilon_y \leq 0.002$ independently. We fixed $\beta_1 = 1.0$, and varied $0.25 \leq \beta_2 \leq 4$. For $0.9 \leq \beta_2 \leq 1.3$, we used $\epsilon_{12} = 1$, but, for $\beta_2 \leq 0.7$ ($\beta_2 \geq 2$), we had to adjust $0.5 \leq \epsilon_{12} \leq 0.75$ ($1.5 \leq \epsilon_{12} \leq 2$) to retain the stability of the heterostructures. We again used $\alpha_{12} = -0.03$ to include the weak oscillations. We determined the average densities $\bar{n}_i^\textrm{(c)}$ and $\bar{n}_i^\textrm{(d)}$ in equilibrium by relaxing bicrystalline heterostructures and by extracting the average densities from the middle of the crystalline phases.

\begin{figure}
    \centering
    \includegraphics[width=0.75\columnwidth]{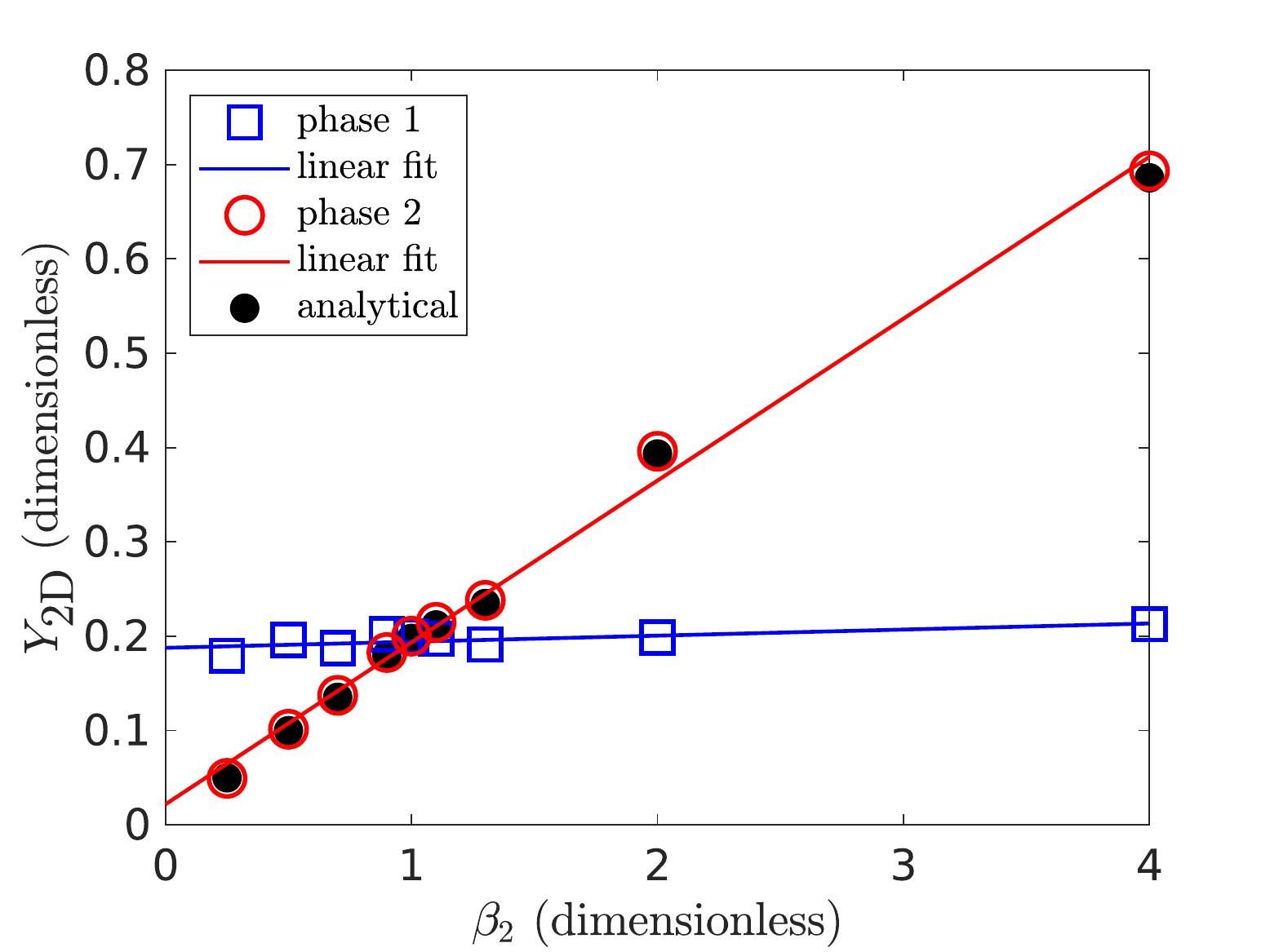}
    \caption{Two-dimensional Young's modulus $Y_\textrm{2D}$ as a function of the gradient term coefficient $\beta_2$ for both crystalline phases. The markers give the actual data and the lines are optimal linear fits. The slope for the second fit is 0.17. The analytical prediction obtained for the crystalline phase 2 using the analytical expressions for the stiffness coefficients $C_{11}$ and $C_{12}$ [Eqs. (\ref{eq-c11}) - (\ref{eq-c12})] is plotted using solid black markers.}
    \label{fig-y-vs-beta}
\end{figure}

Figure \ref{fig-y-vs-beta} shows the 2D Young's modulus as a function of $\beta_2$ for both crystalline phases. For crystalline phase 1, the modulus is essentially unaffected by $\beta_2$, i.e., the corresponding linear fit has a negligible slope. In contrast, the modulus for the crystalline phase 2 is linearly proportional to $\beta_2$ with a slope of 0.17. Independent control of the elastic stiffness of either of the crystalline phases appears straightforward. In addition, for each value of $\beta_2$, we observed $C_{12} \approx C_{11}/3$, whereby $\nu \approx 1/3$. This is a feature common to many simple PFC models \cite{ref-our-prb16}.

Finally, we compared the numerical results against an analytical prediction \cite{ref-pf-book} where for one crystalline phase

\begin{equation}
    \label{eq-c11}
    C_{11} = 9 \sum_i \beta_i \phi_i^2
\end{equation}

\noindent with an amplitude

\begin{align}
    \begin{split}
        \phi_i = & \frac{1}{15 \delta_i} \left( \vphantom{\sqrt{\left( \bar{n}_i \right)}} \gamma_i - 3 \delta_i \bar{n}_i \right.
        \\
        & \left. - \sqrt{\gamma_i^2 - 15 \alpha_i \delta_i + 12 \delta_i \bar{n}_i \left( 2 \gamma_i - 3 \delta_i \bar{n}_i \right)} \right)
    \end{split}
\end{align}

\noindent and

\begin{equation}
    \label{eq-c12}
    C_{12} = C_{11}/3.
\end{equation}

\noindent Here, $C_{11}$ is given simply as a sum over the individual density fields. The values obtained for $Y_\textrm{2D}$ using the analytical expressions above are also plotted in Fig. \ref{fig-y-vs-beta} for the crystalline phase 2. The numerical and analytical results are in very good agreement. Note that the amplitude of the weak oscillations in the crystalline phase 1 is roughly an order of magnitude lower than that of the oscillations in the crystalline phase 2. Since $C_{11} \propto \phi_i^2$, the influence of these oscillations is negligible.

\subsection{Influence of \texorpdfstring{$\nu_i$}{vi}}

\label{sec-nu}

As a final demonstration of our model for binary heterostructures, we introduced lattice mismatch between the two crystalline phases via $\nu_i = 1/\lambda_i$. We fixed $\lambda_1 = 1.0$ and varied $\lambda_2$ through values 1.05, 1.1 and 1.2. We ensured the stability of the heterostructures by choosing $\epsilon_{12} =$ 0.9, 0.75 and 0.5, respectively, and by setting $\alpha_{12} = -$0.03. Here $\beta_i = 1.0$ for simplicity.

We first considered bicrystalline model systems either with different permutations of armchair and zigzag edges along the interface or with symmetrically tilted crystals with a tilt angle $2 \theta = \theta - \left( - \theta \right)$, where $5.5^\circ \leq 2 \theta \leq 55.8^\circ$. 
We considered two different strains. For unstrained systems, the periodicities of both bicrystal halves were matched separately with the periodic domain whereby the lattice mismatch is accommodated by misfit dislocations. For strained systems, both bicrystal halves were initialized with an average lattice constant and were again matched with the domain whereby the lattice mismatch is accommodated via elastic deformation. 
The average densities for the different strain and mismatch cases are given in Table \ref{tab-nu-densities}. Note that said densities were chosen to yield an approximate 1:1 coexistence between the two crystalline phases. For reference, we considered here also $\lambda_2 = 1.0$ with $\epsilon_{12} = 1.0$.

\begin{table}
\centering
\caption{Average densities for lattice-mismatched bicrystalline heterostructures. The mismatch is indicated by $\lambda_2$.}
\label{tab-nu-densities}
\begin{tabular}{ c c c c c }
$\lambda_2$ & $\bar{n}_1^\textrm{(c)}$ & $\bar{n}_1^\textrm{(d)}$ & $\bar{n}_2^\textrm{(c)}$ & $\bar{n}_2^\textrm{(d)}$ \\
\hline
 1.0 & 0.19 & 0.51 & 0.19 & 0.51 \\
 1.05 & 0.16 & 0.55 & 0.12 & 0.55 \\
 1.1 & 0.24 & 0.47 & 0.21 & 0.48 \\
 1.2 & 0.28 & 0.43 & 0.23 & 0.45
\end{tabular}
\end{table}

\begin{figure*}
\centering
    \includegraphics[width=0.75\textwidth]{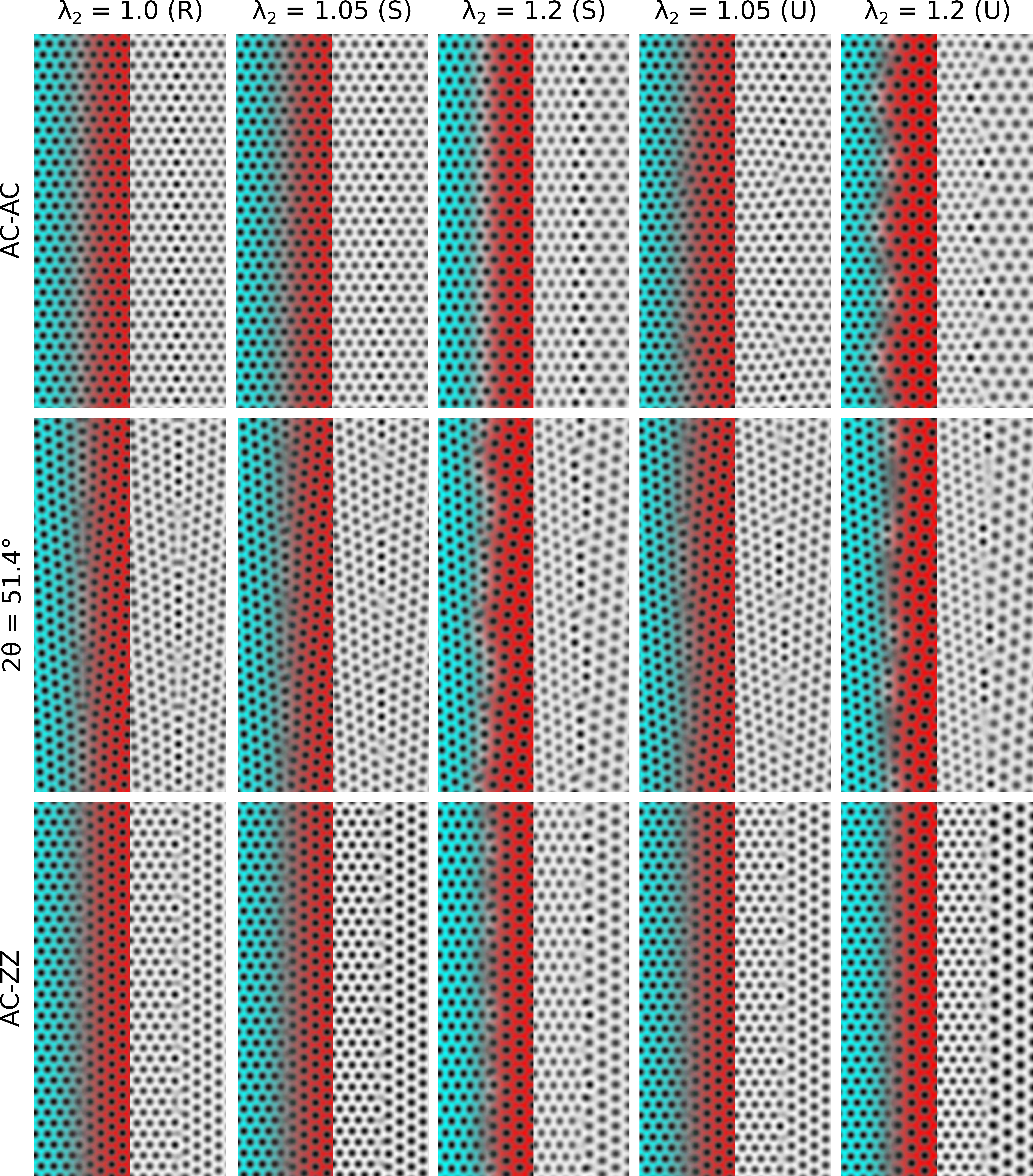}
\caption{Collage of interface structures for different lattice mismatches, strains and misorientations. The left hand side of each panel reveals the distribution of the two crystalline phases, and the right hand side represents the heterostructure by $m = n_1 + n_2$ for a clearer illustration of the atomic level structure. Note that in many cases only a small part of the total length of the interface modeled is shown. The first column of panels gives reference (R) structures with no lattice mismatch between the two phases. The next two columns give strained (S) structures where the mismatch is accommodated by elastic deformation. The last two columns give unstrained (U) structures where the mismatch is accommodated by misfit dislocations. The mismatch for each column is indicated via $\lambda_2 = 1 / \nu_2$. The first row of panels gives structures with zero-misorientation armchair-armchair (AC-AC) interfaces. The second row depicts low-misorientation tilt interfaces with $2 \theta \approx 51.4^\circ$. The third row demonstrates high-misorientation armchair-zigzag (AC-ZZ) interfaces.}
    \label{fig-nu-bi}
\end{figure*}

Overall, the phase interfaces obtained displayed well-defined structures. While misorientation and lattice mismatch introduce defects, extensively fuzzy and ill-defined structures are rare. 
In addition, the vast majority of the highly strained systems remained stable during relaxation. 
Figure \ref{fig-nu-bi} offers a representative sample of the structures obtained and shows a comparison between different strain and mismatch cases.

The first row of panels in Fig.~\ref{fig-nu-bi} demonstrates zero-misorientation armchair-armchair interfaces between lattices of varying mismatch. We observed perfect hexagonal order for the reference and strained cases. Indeed, the strained heterostructures retained their stability without experiencing any stress-relieving reconstructions such as via subsequent nucleation, creation and annihilation of dislocations. In the unstrained structures, we observed periodic arrays of point-like misfit dislocations along the interface. We obtained similar structures for zigzag-zigzag interfaces.

The second row of panels in Fig. \ref{fig-nu-bi} gives low-misorientation interfaces between symmetrically tilted lattices of varying mismatch. Here, the tilt angle $2\theta \approx 51.4^\circ$. All structures display fairly periodic arrays of dislocations. For corresponding graphene grain boundaries, alternatingly slanted dislocations are expected \cite{ref-yazyev-louie, ref-our-prb16}. For the reference structure, the highly symmetric initial state used has lead to extended defect structures without such symmetry breaking. The low-strain structure with $\lambda_2 = 1.05$ indeed displays alternatingly slanted dislocations, whereas the high-strain structure with $\lambda_2 = 1.2$ again does not, due to having achieved some strain-relief via annihilations of dislocations. The unstrained structures appear very similar to the corresponding strained ones.

The third row of panels in Fig. \ref{fig-nu-bi} depicts high-misorientation armchair-zigzag interfaces between lattices of varying mismatch. Despite the extreme misorientation between the two crystalline phases, the present model performs well in stitching them together with rather well-defined atomic-level structures. In fact, the lattice-mismatched structures do not appear visibly fuzzier than the reference.

Finally, we simulated the growth of polycrystalline heterostructures with the aforementioned lattice mismatches. The density fields were initialized with white noise where $\bar{n}_i = [ \bar{n}_i^\textrm{(c)} + \bar{n}_i^\textrm{(d)}] / 2$.

\begin{figure*}
    \centering
    \includegraphics[width=\textwidth]{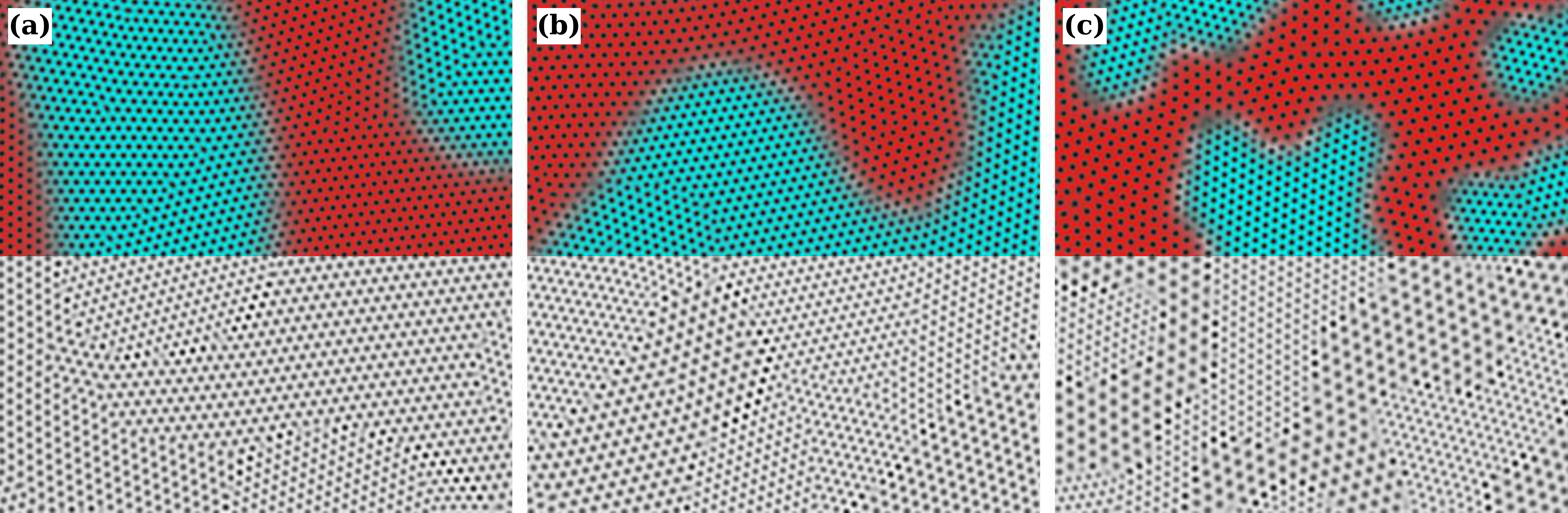}
    \caption{Lattice-mismatched polycrystalline heterostructures. The panels offer blow-ups of larger systems after a relaxation of 250 000 time units. The top halves of the panels show the distribution of the two phases and the bottom halves represent the heterostructures by $m = n_1 + n_2$ for a clearer illustration of the atomic-level structure. We have fixed $\lambda_1 = 1 / \nu_1 = 1$ and have varied $\lambda_2 = 1.05$ in (a), $\lambda_2 = 1.1$ in (b) and $\lambda_2 = 1.2$ in (c).}
    \label{fig-nu-poly}
\end{figure*}

Figure \ref{fig-nu-poly} gives examples of the lattice-mismatched polycrystalline heterostructures obtained. Despite the various misorientations between the two crystalline phases at their interfaces, the model performs well in localizing the mismatch into point-like dislocations. Some interfaces, especially those with larger mismatch, appear somewhat diffuse, but are well comparable to some single-phase PFC grain boundaries; \textit{cf.} Ref. \citenum{ref-atom-based-grain-extraction}, for example. For the cases with $\lambda_2 = 1.05$ and $\lambda_2 = 1.1$, the mismatch is minor and the interfaces appear highly continuous; see panels (a) and (b), respectively. There are a number of dislocations along the interfaces in both heterostructures, but many are due to lattice misorientation. For the case with $\lambda_2 = 1.2$, the interfaces are still fairly continuous, but display several regions with somewhat fuzzy features; see panel (c). These regions seem to coincide with greater interfacial curvature. Last, it also appears that the mobility of the interfaces decreases with increasing mismatch. This is evident from the noticeably smaller domain sizes in the heterostructure with $\lambda_2 = 1.2$; note that all three have been relaxed for the same 250 000 time units.

\section{Application to lateral graphene--hexagonal boron nitride heterostructures}

\label{sec-trinary}

In this section, we consider a three-component model for lateral 2D heterostructures between graphene and h-BN (G--h-BN). 
We focus here on demonstrating the suitability of the model to qualitative modelling of G--h-BN. Finding optimal model parameters for quantitatively accurate modelling of G--h-BN (involving, e.g., fitting to interfacial formation energies) will be presented in future work.

\subsection{Model requirements and parameters}

To model G--h-BN, we set $N = 3$ and chose to model the graphene phase with $n_1^\textrm{(c)}, n_2^\textrm{(d)}$ and $n_3^\textrm{(d)}$ and the h-BN phase with $n_1^\textrm{(d)}, n_2^\textrm{(c)}$ and $n_3^\textrm{(c)}$. The parameters for $n_2, n_3$ and their mutual couplings were adopted from Ref. \citenum{ref-pfc-h-bn}. 
The other parameters were chosen by trial and error by varying them one at a time.
We use the following criteria, guiding principles and simplifying assumptions:

\begin{itemize}
    \item Start by choosing parameters for $n_1$ that are similar to those for $n_2$ and $n_3$.
    \item Assume that boron and nitrogen atoms are interchangeable, i.e., $F \left( n_1, n_2, n_3 \right) = F \left( n_1, n_3, n_2 \right)$.
    \item Match the relative Young's moduli and lattice constants to the experimentally and theoretically determined values $Y_\textrm{h-BN} / Y_\textrm{G} \approx 0.87$ \cite{ref-graphene-modulus, ref-h-bn-modulus} and $a_\textrm{h-BN} / a_\textrm{G} \approx 1.018$ \cite{ref-lattice-constants-1, ref-lattice-constants-2, ref-lattice-constants-3}, respectively.
    \item Require sharp, continuous interfaces and faceted crystal shapes \cite{ref-intermixing, ref-heterostructures-lateral-2, ref-interfaces-3, ref-interfaces-4, ref-interfaces-5, ref-interfaces-6, ref-interfaces-7}.
    \item Retain the stability of the heterostructures and limit the amplitude of weak oscillations.
\end{itemize}

\noindent Table \ref{tab-g-h-bn-parameters} gives a set of model parameters that was found to satisfy the criteria listed above. Most importantly, this choice of parameters yielded $Y_\textrm{h-BN} / Y_\textrm{G} = 0.84$ and $a_\textrm{h-BN} / a_\textrm{G} = 1.021$ in fair agreement with the target values. In the following, we demonstrate how the model behaves and how it fulfills the other criteria above.

\begin{table}
\centering
\caption{Set of parameters for lateral heterostructures of graphene and h-BN. Note that $\alpha_i = \alpha_{ij}$ when $i = j$ and similarly for the other parameters.}
\label{tab-g-h-bn-parameters}
\begin{tabular}{ c c c c }
\hline
\hline
 $N$ &  &  &  \\
 3 &  &  &  \\
\hline
 $\alpha_{ij}$ & $i = 1$ & $i = 2$ & $i = 3$ \\
 $j = 1$ & -1.4     & --    & --    \\
 $j = 2$ & -0.04    & -0.3  & --    \\
 $j = 3$ & -0.04    & 0.5   & -0.3  \\
\hline
 $\beta_{ij}$ & $i = 1$ & $i = 2$ & $i = 3$ \\
 $j = 1$ & 2.25 & --    & --    \\
 $j = 2$ & 0.0  & 1.0   & --    \\
 $j = 3$ & 0.0  & 0.02  & 1.0   \\
\hline
 $\gamma_{ij}$ & $i = 1$ & $i = 2$ & $i = 3$ \\
 $j = 1$ & 0.0  & --    & --    \\
 $j = 2$ & 0.0  & 0.0   & --    \\
 $j = 3$ & 0.0  & 0.3   & 0.0   \\
\hline
 $\delta_{ij}$ & $i = 1$ & $i = 2$ & $i = 3$ \\
 $j = 1$ & 2.25  & --    & --    \\
 $j = 2$ & --   & 1.0   & --    \\
 $j = 3$ & --   & --    & 1.0   \\
\hline
 $\epsilon_{ij}$ & $i = 1$ & $i = 2$ & $i = 3$ \\
 $j = 1$ & --   & --    & --    \\
 $j = 2$ & -0.8 & --    & --    \\
 $j = 3$ & -0.8 & 0.0   & --    \\
\hline
 $\lambda_{ij} = 1 / \nu_{ij}$ & $i = 1$ & $i = 2$ & $i = 3$ \\
 $j = 1$ & 1.0  & --    & --    \\
 $j = 2$ & 0.0  & 1.018 & --    \\
 $j = 3$ & 0.0  & 1.018 & 1.018 \\
\hline
 $\bar{n}_i^{\left( j \right)}$ & $i = 1$ & $i = 2$ & $i = 3$ \\
 $j =$ c & 0.31 & -0.32 & -0.32 \\
 $j =$ d & 0.66 & -0.65 & -0.65 \\
\hline
\hline
\end{tabular}
\end{table}

\subsection{Atomic configurations}

\begin{figure}
    \centering
    \includegraphics[width=0.4\textwidth]{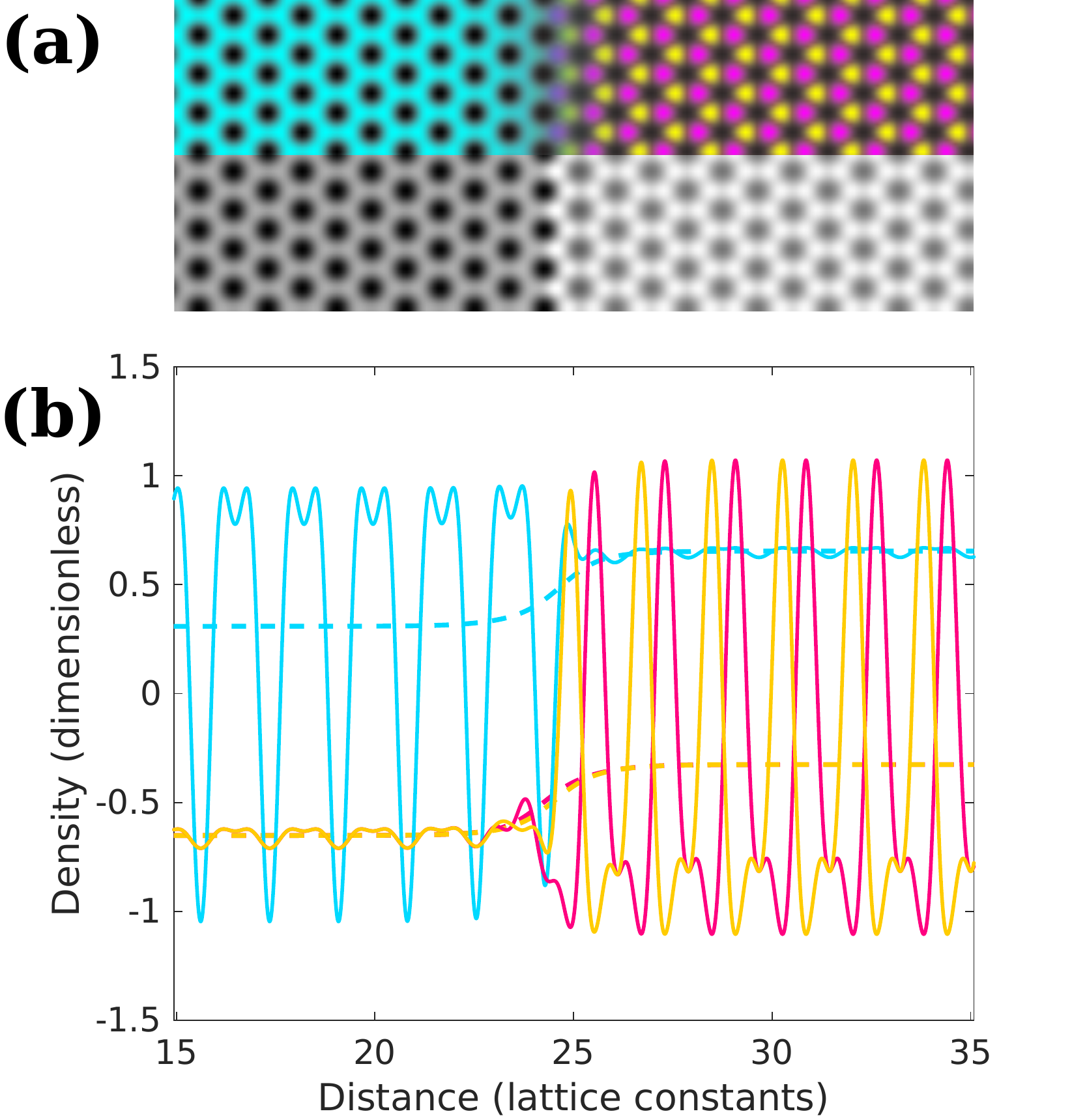}
    \caption{Zigzag interface from a bicrystalline G--h-BN lateral heterostructure. (a) A visualization of the heterostructure in the top half and in the bottom half the same structure representedy by $m = n_1 + n_2 + n_3$ for a clearer illustration of the atomic level structure. In the top half, graphene appears cyan, whereas boron and nitrogen are in magenta and yellow. In the bottom half, graphene (h-BN) appears darker (brighter). (b) Profiles of the densities (solid lines) $n_1$ (cyan), $n_2$ (magenta) and $n_3$ (yellow) and of the smoothed densities (dashed lines) $\eta_1$ (cyan), $\eta_2$ (magenta) and $\eta_3$ (yellow) along the periodic edge in the horizontal direction of the relaxed heterostructure.}
    \label{fig-trinary-bi}
\end{figure}

Figure \ref{fig-trinary-bi} demonstrates a zigzag-oriented interface in a bicrystalline G--h-BN system. Panel (a) gives both the distribution of the two phases and a representation of the same structure by $m = n_1 + n_2 + n_3$ for a clearer illustration of the atomic-level structure. Note that in the latter the h-BN phase appears brighter facilitating identification of the two phases in such figures. The interface displays perfect hexagonal order and is again atomistically sharp. Panel (b) gives the profiles of $n_i$ and $\eta_i$ along the horizontal periodic edge of the system and shows that the amplitude of the weak oscillations is small. Note that in contrast to the binary heterostructures considered in Sec. \ref{sec-binary}, here the average densities $\bar{n}_2$ and $\bar{n}_3$ are negative.

\begin{figure*}
    \centering
    \includegraphics[width=\textwidth]{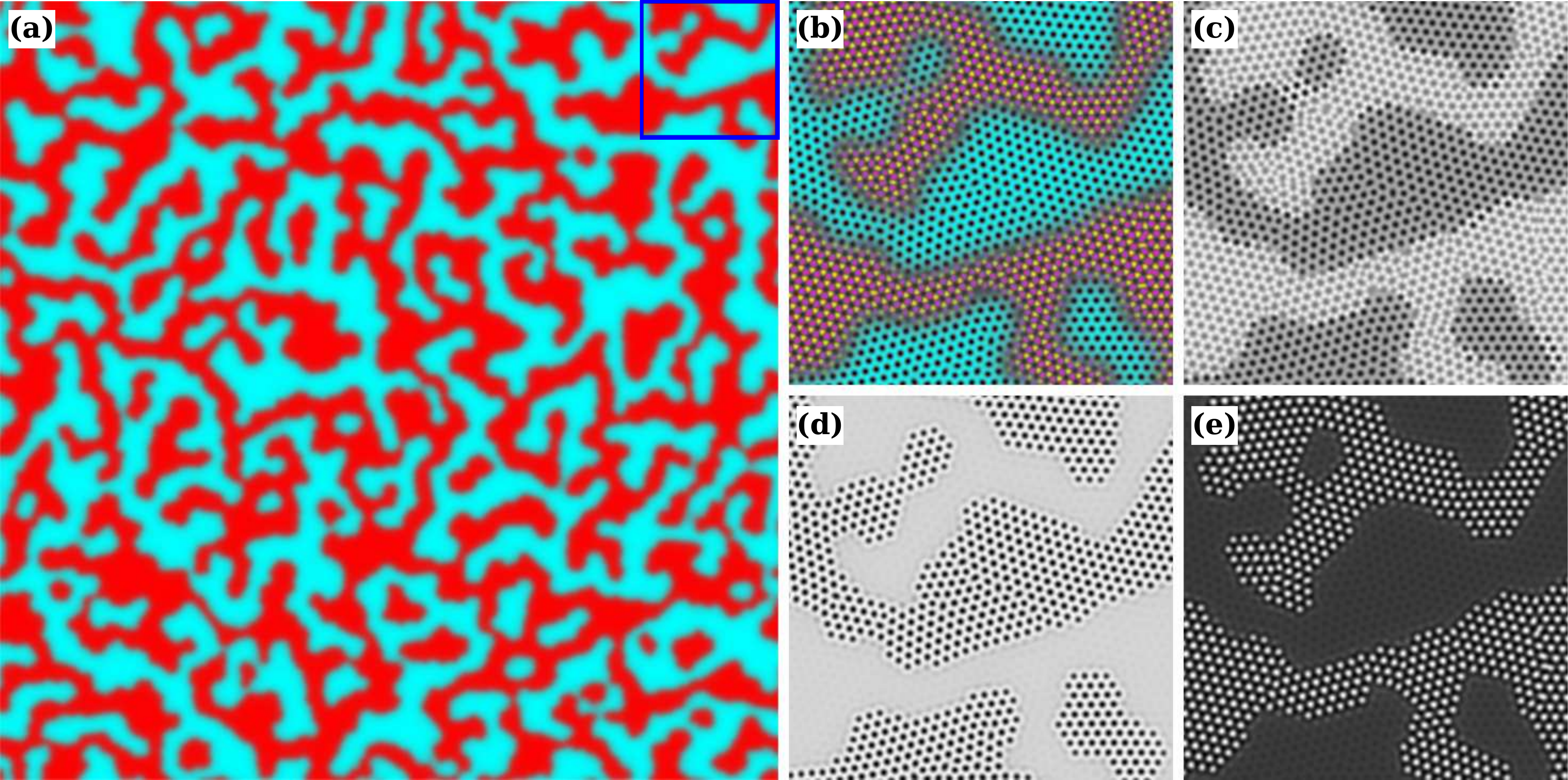}
    \caption{Large random polycrystalline graphene--h-BN lateral heterostructure relaxed from white noise for $2.5 \times 10^6$ time units. The sides of the system are approx. 50 nm in length. (a) A coarse-grained representation of the large-scale structure where graphene appears cyan and h-BN red. (b) - (e) Blow-ups of the region indicated by the blue square in panel (a). The width (height) of the region shown in the blow-ups is approximately 9 nm. (b) A visualization explained in Fig.~\ref{fig-trinary-bi}, (c) the total density $m = n_1 + n_2 + n_3$, (d) $n_1$ and (e) $n_2$ in the blow-up.}
    \label{fig-trinary-poly}
\end{figure*}

Figure \ref{fig-trinary-poly} demonstrates a large polycrystalline G--h-BN system grown from white noise where $\bar{n}_i = [ \bar{n}_i^\textrm{(c)} + \bar{n}_i^\textrm{(d)}] / 2$. In panel (a), an overview of an approx. 50 $\times$ 50 nm$^2$ system is given by a coarse-grained representation where the graphene (h-BN) phase appears cyan (red). After a relaxation of $2.5 \times 10^6$ time units, the heterostructure assumes configurations typical to spinodal decomposition in binary systems. 
Coarsening is slow because we have strived here for stable sharp crystalline structures instead of diffuse high-temperature ones. 
Panels (b) - (e) show a blow-up of the region indicated by a blue square in panel (a). 
Panel (b) visualizes the region and panel (c) presents $m$ for a clearer illustration of the atomic-level structure. 
Despite the various misorientations present in the system, the structure of the interfaces is overall well-defined, excluding few fuzzy patches. Panels (d) and (e) show $n_1$ and $n_2$, respectively. Both appear faceted and have sharp interfaces with a primary (secondary) preference for the zigzag (armchair) direction. Weak oscillations are also visible in both panels.

\begin{figure*}
    \centering
    \includegraphics[width=\textwidth]{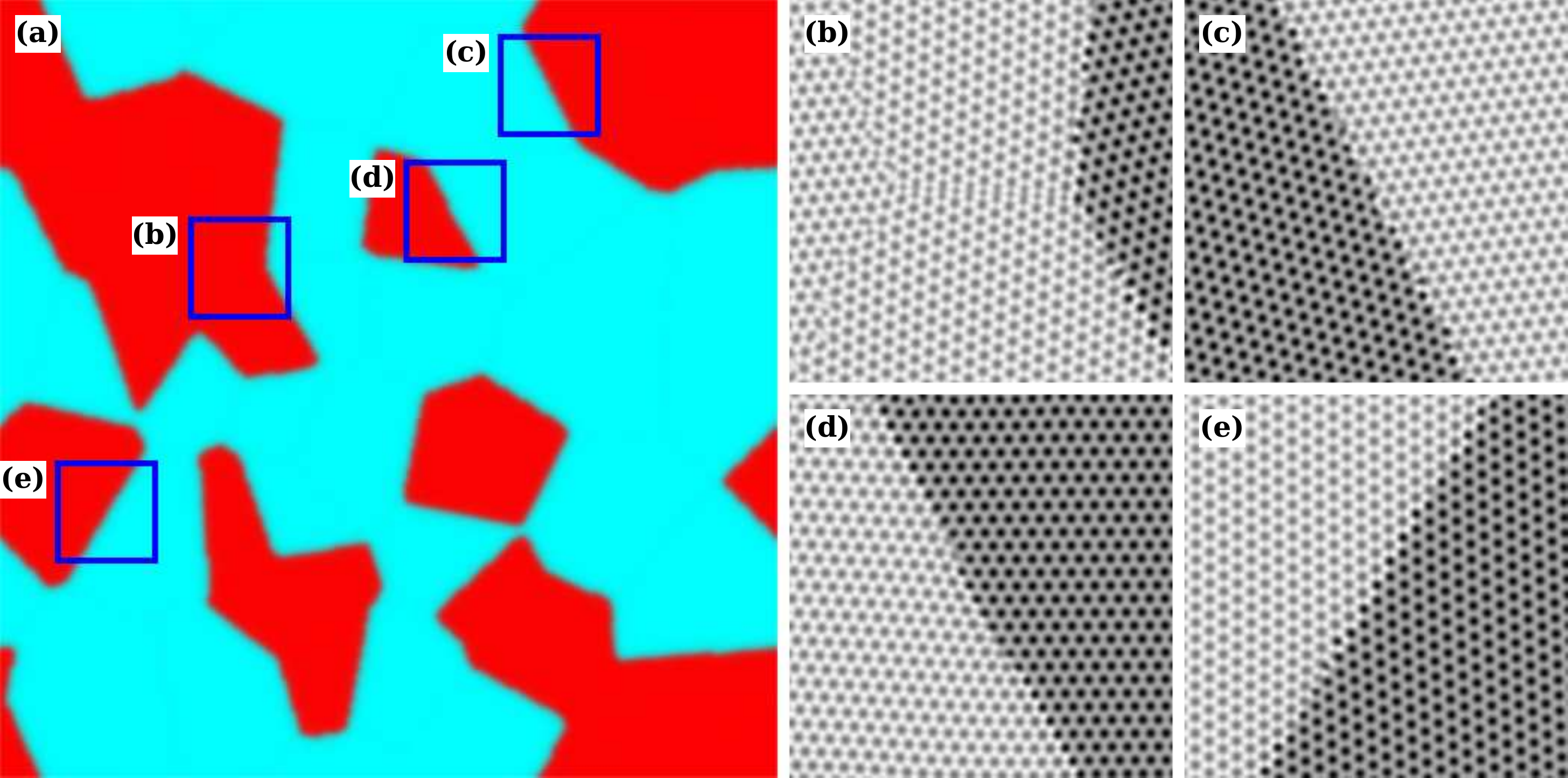}
    \caption{Large random polycrystalline graphene--h-BN lateral heterostructure from a random Voronoi grain structure. A side of the system is approximately 50 nm long. (a) A coarse-grained representation of the large-scale structure where graphene appears cyan and h-BN red. (b) - (e) Blow-ups of the atomic level structure of the regions indicated by the blue squares in panel (a). The regions shown in the blow-ups are 6 nm wide. (b) A collection of graphene--h-BN interfaces and h-BN grain and inversion boundaries. (c) A large-misorientation interface. Small-misorientation (d) zigzag and (e) armchair interfaces.}
    \label{fig-trinary-voronoi}
\end{figure*}

The coarsening of G--h-BN was found slow with the present model and set of parameters. Moreover, concurrent nucleation and growth is not how said heterostructures are produced in practice \cite{ref-heterostructures-lateral-1,ref-heterostructures-lateral-2,ref-heterostructures-lateral-3,ref-heterostructures-lateral-4}. 
We demonstrated preparing more realistic model systems of random polycrystalline G--h-BN with larger phase domain and grain sizes. 
For initialization, we used Voronoi grain structures with random seed points, crystal orientations and phases \cite{ref-our-pccp18} and relaxed for 25 000 time units for local relaxation of the interfaces and grain boundaries. 
Figure \ref{fig-trinary-voronoi} gives an overview of one such system where the initial, large-scale Voronoi structure has remained essentially unchanged as shown by the coarse-grained depiction of the system in panel (a). 
Panels (b) - (e) show the total density $m$ illustrating the atomic level structure of selected interfaces and boundaries. 
Panel (b) displays two triple junctions, one within h-BN and the other between graphene and h-BN, connected by an inversion boundary within h-BN. In h-BN, an inversion boundary is formed between two crystals with a misorientation of $60^\circ$ as the ordering of boron and nitrogen becomes inverted in one crystal with respect to the other \cite{ref-pfc-h-bn, ref-our-pccp18}. 
The interfaces between graphene and h-BN have small-to-intermediate misorientations, whereas the grain boundaries within h-BN are large-angle boundaries. The interfaces and grain boundaries appear disordered but have fairly well-defined atomic level structures. The inversion boundary is formed by a perfect 4|8 chain as expected \cite{ref-h-bn-yakobson, ref-pfc-h-bn}. Panels (c) - (e) show longer interfaces between graphene and h-BN: (c) a large-misorientation interface, (d) a small-misorientation zigzag interface and (e) a small-misorientation armchair interface. While the large-angle interface shown in panel (c) appears disordered, all interfaces display well-defined atomic level structures.

\subsection{Interface energies}

\begin{figure}
    \centering
    \includegraphics[width=\columnwidth]{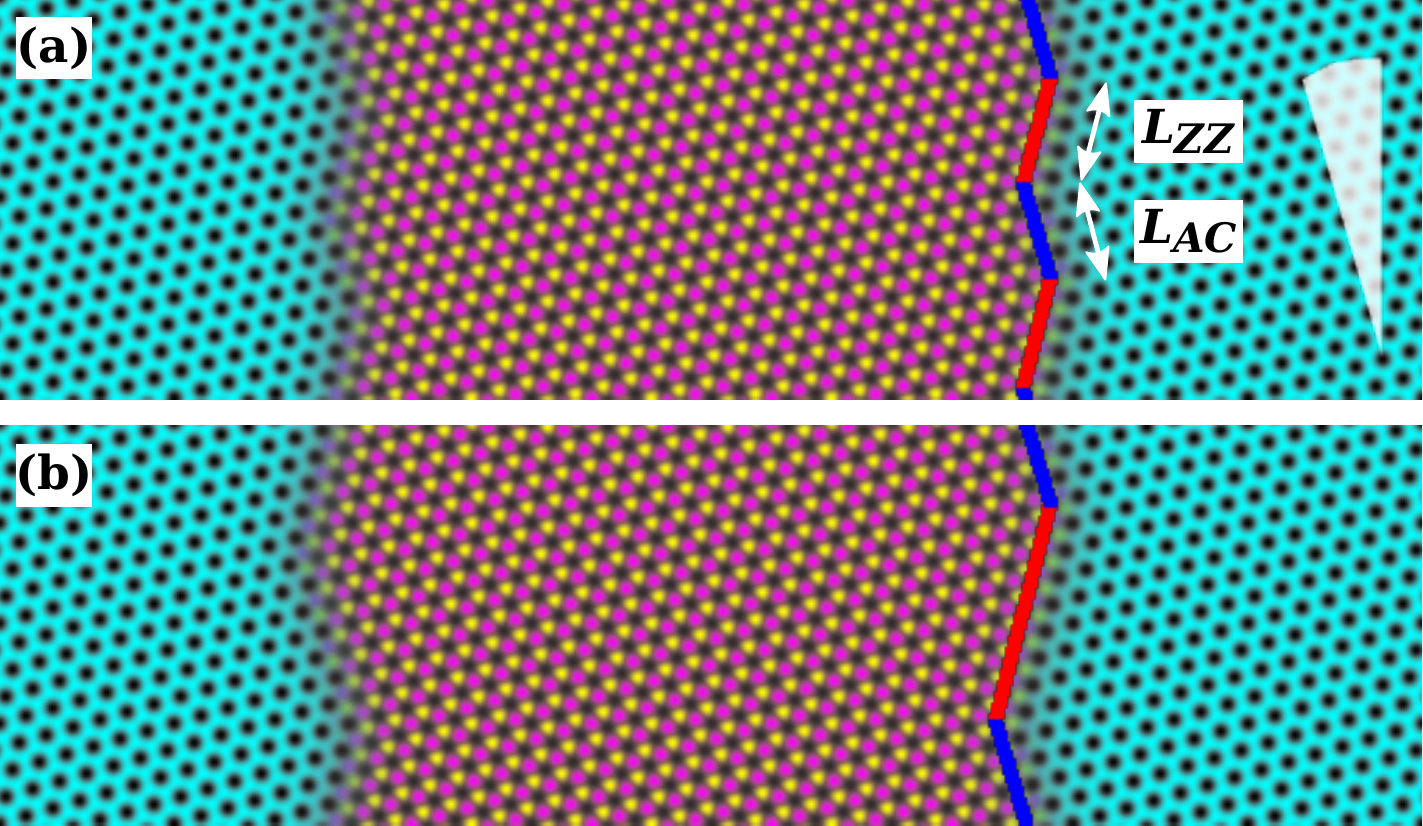}
    \caption{Examples of strained bicrystalline heterostructures used to study the formation energy of G--h-BN interfaces with perfect honeycomb order. The interface angle is $\theta \approx 16.1^\circ$ for both structures and is indicated by the white wedge in panel (a). In panel (a), $m = 2$ and in panel (b) $m = 4$. For both, $n = 1$ and the total width is $L_\bot \approx 12$ nm  (see text for the definition of $m$ and $n$). The zigzag and armchair segments of the right-hand side interfaces are traced in red and blue, respectively. Their lengths $L_{ZZ}$ and $L_{AC}$ are also indicated. In panel (a), $L_{ZZ}$ is 4 lattice constants and $L_{AC}$ is $2 \sqrt{3}$ lattice constants. In panel (b), both are twice as long. Note that the system shown in panel (a) can be decomposed into two identical fields by cutting the domain in half perpendicular to the interface. Here we consider such subdomains with four vertices.}
    \label{fig-bi-interfaces}
\end{figure}

Finally, we investigated the relative stability of G--h-BN interfaces in different lattice directions by studying their formation energies. 
We considered 12 different interface angles $0^\circ \leq \theta \leq 30^\circ$, where $\theta = 0^\circ$ corresponds to armchair and $\theta = 30^\circ$ to zigzag interfaces. 
For simplicity, we considered here only strained configurations with perfect honeycomb order and no misfit dislocations along the interfaces (see Fig. \ref{fig-bi-interfaces} for an example) to avoid very large system sizes. We assumed all interfaces to be composed of zigzag and armchair segments as shown in Fig. \ref{fig-bi-interfaces}. 
The model appears to yield stepped interfaces, typically with minimal segment lengths $L_\textrm{ZZ}$ and $L_\textrm{AC}$. 
Interfaces initialized with longer segments are also at least metastable.

Assuming a sufficiently large bicrystalline heterostructure with dimensions $L_\bot$ perpendicular and $L_\parallel$ parallel to the two interfaces, the total formation energy of the system can be written as

\begin{equation}
    \label{eq-formation}
    F = f L_\perp L_\parallel = f^\ast L_\perp L_\parallel + 2 \gamma L_\parallel + 4 \delta,
\end{equation}

\noindent where $f$ is the free energy density per unit area obtained by evaluating Eq. (\ref{eq-model}) and by dividing by the total area, and $f^\ast = x f_\textrm{G} + \left( 1 - x \right) f_\textrm{h-BN}$ is the effective free energy density per unit area given by the equilibrium bulk free energy densities of the two phases weighted by their area fractions $x$ and $1 - x$. Note that we fixed $x \approx 0.5$ by fixing $\bar{n}_i = \pm 0.485$ and by initializing the two phases with equal or very close to equal areas. Here $\gamma$ is the average formation energy of the interface per unit length (two interfaces appear here; one has C-B and the other C-N bonds along its zigzag segment) and $\delta$ is the average formation energy of a vertex (each system has a total of four vertices; two are convex and the other two concave with respect to one of the two constituents).

The energy terms $f^\ast$, $\gamma$ and $\delta$ can be obtained by fitting Eq.~(\ref{eq-formation}) to simulation data using the method of least squares. For each interface angle considered, we varied $L_\bot = n L_\bot^0$ and $L_\parallel = m L_\parallel^0$, where $L_\bot^0$ is roughly 10 nm, $L_\parallel^0$ is the minimal $L_\parallel$ that satisfies periodic boundary conditions, $n = 1, 2,\dotsc, 5;$  and $m = 1, 2,\dotsc, 8$. For each combination of $L_\bot, L_\parallel$, we initialized a corresponding bicrystalline heterostructure with segmented interfaces, relaxed it, extracted the final $f, L_\bot$ and $L_\parallel$\footnote{The energy $F$ needs to be relaxed with respect to  $L_\bot$ and $L_\parallel$ to relieve possible mechanical stresses}, and fitted Eq. (\ref{eq-formation}) to these data. 
Outliers in the data and visually divergent configurations were excluded from the analysis. 
We determined $\gamma$ and $\delta$ for long segment lengths with large $m$. In practice, we fitted to data points unaffected by nonlinear finite-size effects where typically $n, m > 1$. In addition, we considered minimal segment lengths with $m = 1$ (and $L_\parallel$ hence a constant), \textit{i.e.}, interfaces with the maximal packing density of vertices. In this case the vertex energy needs to be absorbed into the interface energy as they cannot be separated without varying $L_\parallel$. This gives the scaling relation

\begin{equation}
    \label{eq-formation-minimal}
    f L_\bot = f^\ast L_\bot + 2 \gamma^\ast,
\end{equation}
where $\gamma^\ast = \gamma + 2 \delta/L_\parallel$. This scaling relation is one-dimensional. Note that for long segment lengths the energy contribution of the vertices is negligible $\gamma \gg 2 \delta / L_\parallel$, and hence $\gamma^\ast \approx \gamma$.

In the limit of long segment lengths, one can derive an analytical expression for $\gamma$ as a sum of the two segments' individual contributions. Since the angle between the zigzag and armchair segments is 150$^\circ$, and as we know $\theta$ and $L_\parallel$, simple geometrical considerations give \cite{ref-graphene-segments}

\begin{equation}
    \label{eq-gamma-analytical}
    \gamma = 2 \gamma_\textrm{ZZ} \sin{\left( \theta \right)} + 2 \gamma_\textrm{AC} \sin{\left( 30^\circ - \theta \right)},
\end{equation}

\noindent where $\gamma_\textrm{ZZ} = \gamma \left( \theta = 30^\circ \right)$ and  $\gamma_\textrm{AC} = \gamma \left( \theta = 0^\circ \right)$ are the interface energies of pure zigzag and armchair interfaces, respectively.

\begin{figure}
    \centering
    \includegraphics[width=\columnwidth]{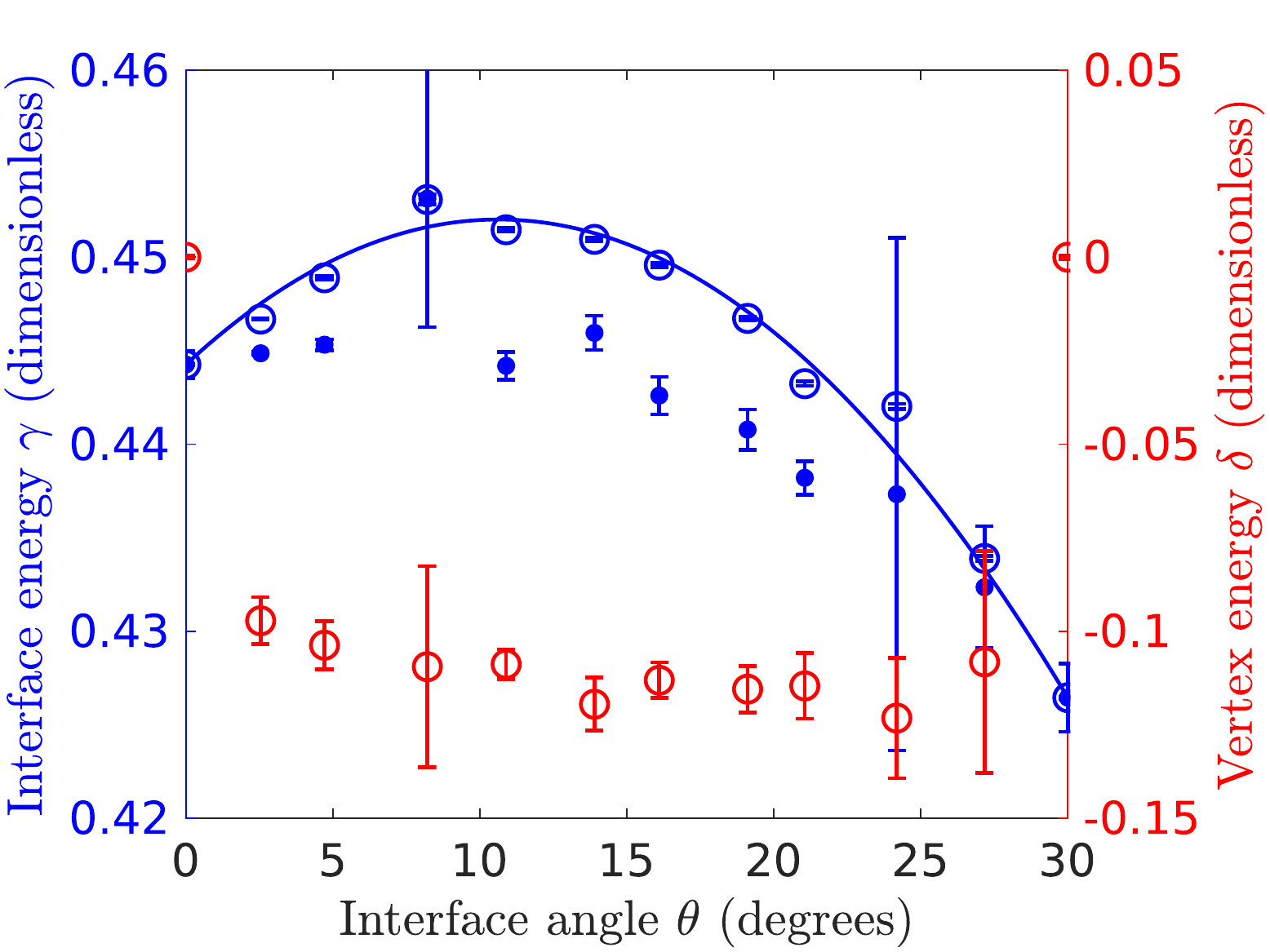}
    \caption{Dimensionless interface $\gamma$ (in blue on left axis) and vertex $\delta$ (in red on right axis) energies as a function of the interface angle $\theta$; see Eq.~(\ref{eq-formation}). Open markers correspond to $\gamma$ in the limit of long segment lengths (large $m$) and solid markers to $\gamma$ in the limit of minimal segment lengths ($m = 1$); see Eq. (\ref{eq-formation-minimal}). The solid curve gives the analytical expression for $\gamma$ from Eq. (\ref{eq-gamma-analytical}).
    }
    \label{fig-gamma-delta}
\end{figure}

Figure \ref{fig-gamma-delta} shows the dimensionless interface $\gamma$ and vertex energies $\delta$  as a function of the interface angle $\theta$. All error bars are given by two-sigma confidence intervals.
In the limit of long segment lengths, the numerical data for $\gamma$ agrees well with the analytical expression given by Eq.~(\ref{eq-gamma-analytical}) predicting a maximal interface energy at $\theta \approx 10^\circ$. One should note that both the zigzag and armchair interfaces have a locally minimal interface energy with respect to the interface angle. In this case the zigzag interface has the lower energy. This is consistent with a number of experimental findings \cite{ref-interfaces-3, ref-interfaces-7, ref-zigzag, ref-intermixing, ref-interfaces-6, ref-interfaces-5}, reporting a preference towards zigzag interfaces. 
The dominance of zigzag interfaces can be explained by the growth process: zigzag-faceted crystals of one phase are typically formed first and serve as seeds for the subsequent growth of the second phase. 
The thermodynamic stability of the zigzag interfaces has also been verified computationally using density functional (DFT) theory \cite{ref-interfaces-3, ref-interfaces-7, ref-interfaces-8}. A contradictory preference for armchair interfaces has also been reported by some DFT studies \cite{ref-armchair-1, ref-armchair-2}.

Figure \ref{fig-gamma-delta} also gives the interface energy for interfaces with minimal segment lengths ($m = 1$), i.e., interfaces with the maximal packing density of vertices. Again, the zigzag direction yields the lowest energy. However, here $\gamma$ is approximately constant for $\theta \leq 15^\circ$ and decreases with $\theta$ for $\theta \geq 30^\circ$. There are two data points with significantly larger error bars. For these cases, the scaling is not as linear as for the other interface angles despite visually ideal configurations. It appears that interfaces with minimal segment lengths have lower formation energies in general.

Figure \ref{fig-gamma-delta} also shows that the vertex energy $\delta$ is zero for both zigzag and armchair directions where there are no vertices. For the intermediate directions, $\delta$ is found slightly negative and roughly constant. This explains why the formation energy is generally lower for interfaces with more vertices. A negative $\delta$ is possible, since the vertices cannot exist independently of the segments whose $\gamma > 0$. Furthermore, interfaces with long segment lengths proved stable as the highly symmetric initial states provided insufficient driving force to overcome the energy barriers for nucleating more vertices. Related point defects, triple junctions between grain boundaries, have also been shown to display negative formation energies \cite{ref-tjs-king, ref-tjs-srinivasan, ref-our-sr17}.

\section{Summary and outlook}

\label{sec-conclusions}

We have introduced an efficient and flexible phase field crystal model intended for studying heterostructures or composite materials for the general case of $N$ atomic species. This model allows well-controlled phase separation via the use of smoothed density fields. The lattice symmetries and the length scale, as well as the elastic properties, of the individual phases can be controlled readily. This model offers a straightforward approach to modelling systems with multiple ordered phases.

We have carried out a comprehensive demonstration of the model's properties using simple 2D binary systems. More specifically, we have varied several model parameters independently to investigate their influence on phase separation, on the atomic-level structure and order of the phase interfaces and on the elastic properties of the two phases. A lattice constant mismatch between the two phases results in disordered but generally well-connected interfacial configurations. The elastic properties of the two phases can be controlled independently and robustly. We have also demonstrated the applicability of this model by considering graphene--hexagonal boron nitride lateral heterostructures (G--h-BN). We have shown that the model can reproduce many of the features of G--h-BN, such as the relative lattice constants and Young's moduli of the two phases, as well as continuous interfaces with a preference for zigzag and armchair directions. We have also demonstrated how to model large, complex G--h-BN microstructures.

One obvious extension to this study would be to further optimize the model parameters used for G--h-BN. While we have matched the relative lattice constants and Young's moduli approximately to their experimental values, most model parameters were either adopted from previous works or were chosen on qualitative grounds. Especially the coupling coefficients in the model could be fitted by matching the structure and formation energy of phase interfaces to corresponding results from atomistic calculations. The different chemical affinities between carbon and boron, and carbon and nitrogen, in particular, could be incorporated to the model via these parameters. In addition, structural or other more sophisticated PFC models could be incorporated to the model by replacing the terms in the energy $F$ proportional to $\beta_i$ and $\beta_{ij}$ with convolution kernels for more accurate material description or to allow a broader range of lattice symmetries. Although here we have focused on the 2D case for conceptual simplicity, the model is also applicable to 3D problems, where various nanoscale heterostructures and mesoscopic multiphase microstructures or composite materials could be considered. Constraining the coupling between the different lattices to their mutual interfaces could facilitate eliminating the occasional fuzzy structures without amplifying the weak oscillations. Achieving this without making the equations of motion significantly more complicated is a topic for future work.

\section{Acknowledgments}

This work has been supported in part by the Academy of Finland through its QFT Center of Excellence Program grant (no. 312298). We acknowledge the computational resources provided by the Aalto Science-IT project and the CSC IT Center for Science, Finland. P.H. acknowledges financial support from the Vilho, Yrj\"o and Kalle V\"ais\"al\"a Foundation of the Finnish Academy of Science and Letters. 
 K.R.E. acknowledges financial support from the
National Science Foundation under Grant No. DMR-1506634 and from the Aalto Science Institute (ASCI).

\bibliography{main}

\appendix

\section{Phase separation}

\label{sec-appendix-a}
\begin{figure}
	\label{fig:liquid_solid_areas}
	\includegraphics[width = 3in]{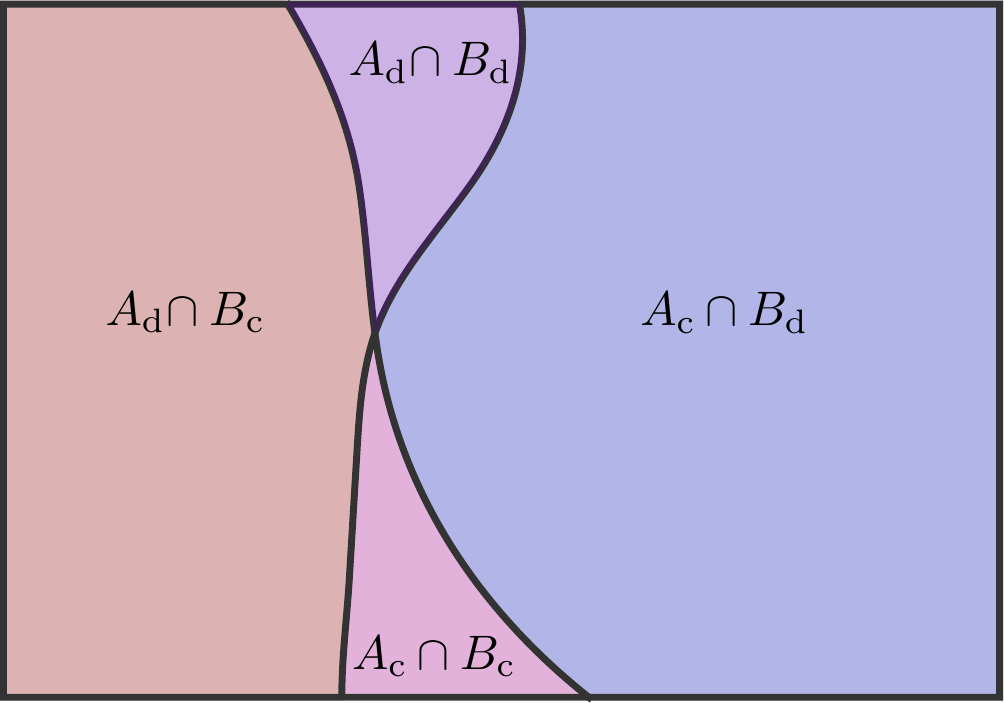}
	\caption{A pictorial showing an example of disordered and ordered regions of the two components A and B. As a consequence of $ V(A_\text{d}) = V(B_\text{c}) $ (and $ V(A_\text{c}) = V(B_\text{d}) $),
		the areas $ V(A_\text{c} \cap B_\text{c}) $ and $ V(A_\text{d} \cap B_\text{d}) $ are equal. See text for details.
	}
\end{figure}

In this Appendix we examine the smoothed phase separating part of the energy 
\begin{equation}
\label{eq:cross_term}
F_{\text{ct}} = \int_\Omega \diff \vec{r} \left[
\epsilon_{\rm AB} \eta_{A}(\vec{r}) \eta_{B} (\vec{r})
\right]
\end{equation}
appearing in Eq.~\eqref{eq-model}. 
Here $ \Omega $ is the domain that can be thought of as a box with a finite volume and periodic boundary conditions (flat torus). 
We neglect the contribution of the disorder--crystalline boundary and assume that $ \eta_{A} $ and $ \eta_{B} $ take constant values specific to the phase. 
Let
\begin{equation}
X_i = \lbrace 
\vec{r} \in \Omega : \eta_X(\vec{r}) = \eta_{X}^{(i)}
\rbrace,
\end{equation}
where $ X $ is the component label $ A $ or $ B $ and $ i \in \lbrace \text{d}, \text{c} \rbrace $ corresponding to \emph{disordered} and \emph{crystalline} phases. We define $ V(X) $ as the volume (area) of set $ X $. A diagram of the setup is shown in Fig.~\ref{fig:liquid_solid_areas}. 
The system is set up such 
that $ V(A_\text{d}) = V(B_\text{c}) $.
Since $ X_\text{d} $ and $ X_\text{c} $ split $ \Omega $ perfectly, also $ V(A_\text{c}) = V(B_\text{d}) $. 
This can be done by choosing the overall number of constituents $ A $ and $ B $ correctly.

Now the energy contribution from Eq.~\eqref{eq:cross_term} is 
\begin{widetext}
\begin{equation}
\begin{split}
	F_{\text{ct}} &=\epsilon_{\rm AB} \left[
	\int_{A_\text{c} \cap B_\text{c} } \diff \vec{r}\, \eta_{A}^\text{(c)} \eta_{B}^\text{(c)}+
	\int_{A_\text{c} \cap B_\text{d} } \diff \vec{r}\, \eta_{A}^\text{(c)} \eta_{B}^\text{(d)}+
	\int_{A_\text{d} \cap B_\text{c} } \diff \vec{r}\, \eta_{A}^\text{(d)} \eta_{B}^\text{(c)}+
	\int_{A_\text{d} \cap B_\text{d} } \diff \vec{r}\, \eta_{A}^\text{(d)} \eta_{B}^\text{(d)}
	\right] \\
	&= \epsilon_{\rm AB} \left[
	V(A_\text{c} \cap B_\text{c})  \eta_{A}^\text{(c)} \eta_{B}^\text{(c)}+
	V(A_\text{c} \cap B_\text{d})  \eta_{A}^\text{(c)} \eta_{B}^\text{(d)}+
	V(A_\text{d} \cap B_\text{c})  \eta_{A}^\text{(d)} \eta_{B}^\text{(c)}+
	V(A_\text{d} \cap B_\text{d})  \eta_{A}^\text{(d)} \eta_{B}^\text{(d)}
	\right]. 
\end{split}
\end{equation}
\end{widetext}

Any set $ Y \subset \Omega$ can be divided such that
 \[  V(Y) = V(Y \cap X_l) + V(Y \cap X_s)  \] 
because $ X_l $ and $ X_s $ split $ \Omega $. 
From this it follows that 
\[ V(A_\text{c} \cap B_\text{d}) = V(A_\text{c}) - V(A_\text{c} \cap B_\text{c})  \] 
and 
\[ V(A_\text{d} \cap B_\text{c}) = V(B_\text{c}) - V(A_\text{c} \cap B_\text{c}).  \]
Moreover, 
\[ 
\begin{split}
V(A_\text{c} \cap B_\text{d}) + V(A_\text{c} \cap B_\text{c}) 
&=  V(A_\text{c}) = V(B_\text{d}) \\
&= V(A_\text{d} \cap B_\text{d}) + V(A_\text{c} \cap B_\text{d}),
\end{split}
\] 
from which it follows that  
\[ V(A_\text{d} \cap B_\text{d}) = V(A_\text{c} \cap B_\text{c}).  \]

Now
\begin{widetext}
	\begin{equation}
\begin{split}
	F_{\text{ct}} &= \epsilon_{\rm AB} \left[
	V(A_\text{c} \cap B_\text{c}) (\eta_{A}^\text{(c)} \eta_{B}^\text{(c)} + \eta_{A}^\text{(d)} \eta_{B}^\text{(d)} - \eta_{A}^\text{(c)} \eta_{B}^\text{(d)} - \eta_{A}^\text{(d)} \eta_{B}^\text{(c)} ) 
	+ V(A_\text{c}) \eta_{A}^\text{(c)} \eta_{B}^\text{(d)} 
	+ V(B_\text{c}) \eta_{A}^\text{(d)} \eta_{B}^\text{(c)}
	\right] \\
	&= \epsilon_{\rm AB} \left[
	V(A_\text{c} \cap B_\text{c}) 
	(\eta_{A}^\text{(c)} - \eta_{A}^\text{(d)}) (\eta_{B}^\text{(c)} - \eta_{B}^\text{(d)})
	+ V(A_\text{c}) \eta_{A}^\text{(c)} \eta_{B}^\text{(d)} 
	+ V(B_\text{c}) \eta_{A}^\text{(d)} \eta_{B}^\text{(c)}
	\right].
\end{split}
\end{equation}
\end{widetext}
The components $ A $ and   $ B $ have a similar phase diagram in the sense that either 
\[ 
\eta_{A}^\text{(c)} > \eta_{A}^\text{(d)}, \; \eta_{B}^\text{(c)} > \eta_{B}^\text{(d)}
 \]
or 
\[ 
\eta_{A}^\text{(c)} < \eta_{A}^\text{(d)}, \; \eta_{B}^\text{(c)} < \eta_{B}^\text{(d)}.
 \]
From this it follows that $ (\eta_{A}^\text{(c)} - \eta_{A}^\text{(d)}) (\eta_{B}^\text{(c)} - \eta_{B}^\text{(d)}) > 0 $, which implies that 
\begin{equation}
	F_{ct} \geq \epsilon_{\rm AB} \left[
	V(A_\text{c}) \eta_{A}^\text{(c)} \eta_{B}^\text{(d)} 
	+ V(B_\text{c}) \eta_{A}^\text{(d)} \eta_{B}^\text{(c)}
	\right].
\end{equation}
Therefore, at the ground state $ V(A_\text{c} \cap B_\text{c})  = 0  $. This shows that $ F $ is minimized when different phases of the different components appear together. Even if $ V (A_\text{c})  \neq V (B_\text{d})$, the areas $ V (A_\text{c}  \cap  B_\text{c}) $ and  $ V (A_\text{d} \cap  B_\text{d}) $ would be minimized. If $ \epsilon_{\rm AB} < 0 $, $ V(A_\text{c} \cap B_\text{c})  = 0  $ is maximized and the crystalline phases of the constituents overlap.

\section{Elastic effects due to smoothed density fields }

\label{sec-appendix-b}
In this Appendix we will study the non-local effects due to the introduction of the smoothed number density fields $ \eta_{\rm X} $ ($ {\rm X} \in \{ {\rm A, B} \} $ )  in $ F $ (Eq.~\eqref{eq-model} ) 
The smoothed number densities appear in the term
\begin{equation}
\label{eq:smoothing_energy}
F_{\text{ct}} = \int \diff \vec{x} \left[
\epsilon_{\rm AB} \eta_{\rm A}(\vec{x}) \eta_{\rm B} (\vec{x})
\right]
\end{equation}
that might contribute to excess elastic energy if the system is deformed. Throughout this section we assume $ \nu_{i} = 1 $. This sets the length scale of the bulk oscillations of the density fields.

The smoothed fields are defined as
\begin{equation}
\label{eq:eta_definition}
\eta_{\rm X}(\vec{x}) = \int \diff \vec{y} \left[
G(\vec{x} - \vec{y}) n_{\rm X}(\vec{y})
\right].
\end{equation}
This convolution gives rise to non-local self-interactions. Let
\begin{equation}
\hat{f}(\vec{k}) = \int \diff \vec{x} \left[
e^{-i \vec{k} \cdot \vec{x}} f(\vec{x})
\right]
\end{equation}
be the Fourier transform of $ f(\vec{x}) $. Now the inverse 2D transform is 
\[  
f(\vec{x}) =  \frac{1}{4\pi^2} \int \diff \vec{x} \left[
e^{i \vec{k} \cdot \vec{x}} \hat{f}(\vec{x})
\right].
\]

The energy $ F_{\rm ct} $ can be written in terms of the Fourier transforms as 
\begin{equation}
F_{\rm ct} = \frac{1}{4 \pi^2} \int \diff \vec{k} \left[
\epsilon_{\rm AB} \hat{\eta}_{\rm A}^*(\vec{k}) \hat{\eta}_{\rm B}(\vec{k})
\right]
\end{equation}
by using the Plancherel theorem. Here $ \hat{\eta}_{\rm A}^* $ is the complex conjugate of $ \hat{\eta}_{\rm A} $. The fields $ \hat{\eta}_{\rm X} $ ($ \rm X \in \{ \rm A, \rm B  \} $ ) can be easily expressed using the convolution theorem as 
\begin{equation}
\hat{\eta}_{\rm X}  = \hat{G} \hat{n}_{\rm X},
\end{equation} \\
where $ \hat{G} $ is the Fourier transform of the Gaussian convolution kernel, also a Gaussian
\begin{equation}
\hat{G}(\vec{k}) = e^{-(\gamma k)^2}.
\end{equation}
Here $ k = |\vec{k}| $ and $ \gamma $ gives the length scale of the smoothing. Notice that $ \hat{G} $ is real. Now 
\begin{equation}
F_{\rm ct} = \frac{\epsilon_{\rm AB}}{4 \pi^2} \int \diff \vec{k} \left[
\hat{G}(\vec{k})^2 \hat{n}_{\rm A}^*(\vec{k}) \hat{n}_{\rm B}(\vec{k})
\right].
\end{equation}

We consider deformations of the form $ \vec{k} - \vec{p} (\vec{k})$, with $ | \vec{p}| \ll |\vec{k}| $. As an example, for a uniform compression of 5\% ,  $\vec{p} \approx 0.05 \vec{k} $. Now $ F_{\rm ct} $ becomes
\begin{equation*}
F_{\rm ct} = \frac{\epsilon_{\rm AB}}{4 \pi^2} \int \diff \vec{k} \left[
\hat{G}(\vec{k})^2 \hat{n}_{\rm A}^*(\vec{k}-\vec{p}) \hat{n}_{\rm B}(\vec{k}-\vec{p})
\right] .
\end{equation*}
Making a change of variables $ \vec{k} \to \vec{k} + \vec{p} $ gives
\begin{equation}
F_{\rm ct} = \frac{\epsilon_{\rm AB}}{4 \pi^2} \int \diff \vec{k}\, \nu(\vec{k})  \left[
\hat{G}(\vec{k} +\vec{p})^2 \hat{n}_{\rm A}^*(\vec{k}) \hat{n}_{\rm B}(\vec{k})
\right],
\end{equation}
where $ \nu $ is the change in the volume element that is given by the determinant of the Jacobian $ \mathbf{I} + \nabla \vec{p} $. 

The fields $ \hat{n}_{\rm X} $ have non-zero structure at the nearest neighbor length scale (PFC fluctuations) and close to $ \vec{k} = 0 $ (order--disorder boundaries). 
The length scale given by $ 1/k = 1$ corresponds to nearest neighbor distance of the PFC lattice and $ \gamma $ is chosen such that $ \hat{G}(k=1) \ll 1 $ implying that deformations at this length scale do not contribute to $ F_{\rm ct} $. 
We will investigate the other important regime, where $ \vec{k} $ is small. 
Let $F_{\rm ct} = \int \diff \vec{k}\, f_{\rm ct} .$ 
Expanding $ \hat{G} $ around $ \vec{k} $ gives
\begin{equation*}
\begin{split}
f_{\rm ct} &\approx \frac{\epsilon_{\rm AB}}{4  \pi^2}  \nu  \hat{n}_{\rm A}^*\hat{n}_{\rm B} \left[
1 + \frac{1}{2} \sum_{i,j} \delta_i \delta_j \partial_{ij} 
\right]\hat{G}(\vec{k})^2 \\
& \approx \frac{\epsilon_{\rm AB}}{4 \pi^2}  \hat{n}_{\rm A}^* \hat{n}_{\rm B}  \left[
1  +  8 (\vec{p} \cdot \vec{k})^2 \gamma^4  - 2 |\vec{p}|^2 \gamma^2
\right] \hat{G}(\vec{k})^2.
\end{split}
\end{equation*}
Here we have used the fact that for small $\vec{p}$, $ \nu \approx 1+\nabla \cdot \vec{p} $ and assume that the part proportional to $\nabla \cdot \vec{p}$ is much smaller than unity. Also, the system is initially in equilibrium implying that $\hat{G}(\vec{k}+ \vec{p})^2$ has to be expanded up to second order. 
Now the excess part of the energy at $\vec{k}$ is
\begin{equation}\label{eq:excess_elastic_energy}
\begin{split}
\Delta f_{\rm ct} &:= f_{\rm ct} - \left. f_{\rm ct} \right|_{\vec{p} = 0} \\
&= \frac{ \epsilon_{\rm AB} \gamma^2}{ 2 \pi^2}   \left[
4 \gamma^2 (\vec{p} \cdot \vec{k})^2  -  |\vec{p}|^2 
\right]  \hat{G}(\vec{k})^2 \hat{n}_{\rm A}^* \hat{n}_{\rm B}.
\end{split}
\end{equation}

In order to calculate the contribution to the excess elastic energy due to an interface, we assume that $ \hat{n}_{\rm A} $ and $ \hat{n}_{\rm B} $ vary only in one direction and are peaked around $ \vec{k} = 0 $. Furthermore, we can estimate $ \hat{n}_{\rm A}^* \hat{n}_{\rm B} < \phi^2$, where $ \phi $ is the amplitude of the one-mode oscillations in the crystal. More precisely
\[ 
\hat{n}_{\rm A}^*(\vec{k}) \hat{n}_{\rm B} (\vec{k}) = 2\pi \delta(k_y) \hat{n}_{\rm A}^*(k_x) \hat{n}_{\rm B} (k_x) (\vec{k}) < 2\pi \delta(k_y) \phi^2.
 \]
The Fourier amplitudes due to the interfaces should be significantly smaller than the Fourier amplitude of the bulk oscillations. The excess energy $ \Delta f_{\rm ct} $ is maximized for parallel $ \vec{k} $ and $ \vec{p} $. Let us assume that $ \vec{p}(\vec{k}) < \delta k_x   $, where $ k_x $ is the component of $ \vec{k} $ parallel to the interface and $ \delta $ is small. 
Now we can estimate the contribution of the interface per interface length as
\begin{equation}\label{eq:interface_excess_elastic_energy}
\begin{split}
\Delta  f_{\rm ct}^{\text{int}} &< \frac{ \epsilon_{\rm AB} \delta^2 \gamma^2 \phi^2}{ \pi}  \int_{-\infty}^{\infty} \diff k_x \left[
4 \gamma^2 k_x^4  -  k_x^2 
\right]  \hat{G}(k_x)^2 \\ 
&= \frac{\epsilon_{\rm AB} \delta^2 \phi^2 }{2\sqrt{2 \pi} \gamma}.
\end{split}
\end{equation} 

We can compare $ \Delta  f_{\rm ct}^{\text{int}} $ to the bulk elastic excess energy. The bulk elastic energy density due to component $ A $ is
\begin{equation}\label{eq:bulk_elastic_energy}
	f_{\rm el}  = \lim\limits_{V_{\Omega} \to \infty} \frac{1}{V_{\Omega}} \int_{\Omega} \diff \vec{x} 
	\left[
	\frac{\beta_{\rm A}}{2} n_{\rm A} (1 + \nabla^2) n_{\rm A}
	\right],
\end{equation}
where $ \Omega $ is a compact domain that is taken to infinity and $ V_{\Omega} $ is its area. We consider the contribution of the bulk oscillations and set 
\begin{equation}\label{eq:bulk_oscillations_raw}
n_{\rm A} = \phi \sum_{j} e^{i \vec{q}_j \cdot \vec{x}},
\end{equation}
where $ \vec{q}_{j} $ are the principal reciprocal lattice vectors with $ q_{j} = 1 $. Now 
\begin{equation}\label{eq:bulk_oscillations_Fourier}
\hat{n}_{\rm A} = 2\pi \phi \sum_{j} \delta(\vec{k} - \vec{q}_{j}).
\end{equation}

We evaluate $ f_{\rm el} $ in Fourier space as 
\begin{equation}\label{eq:bulk_elastic_energy_Fourier}
f_{\rm el}  = \lim\limits_{V_{\Omega} \to \infty} \frac{\beta_{\rm A}}{8 \pi^2 V_{\Omega}} \int \diff \vec{k} |\hat{n}_{\rm A}|^2 \hat{\mathcal{L}}(\vec{k})^2,
\end{equation}
where 
\begin{equation}
\mathcal{L}(\vec{k}) = (1 - k^2)
\end{equation}
is the Fourier transform of the pattern forming operator $ (1 + \nabla^2) $. We repeat the earlier calculation by replacing $ \hat{G} $ with $ \hat{\mathcal{L}} $. We use
\[ 
\lim\limits_{V_{\Omega} \to \infty } \frac{\delta(\vec{k} - \vec{q}_{j}) \delta(\vec{k} - \vec{q}_{i})}{V_{\Omega}} = \delta_{ij} \delta(\vec{k} - \vec{q}_{j}).
 \]
Now
\begin{equation}\label{key}
f_{\rm el} =  \frac{\beta_{\rm A} \phi^2}{2} \int \diff \vec{k} \mathcal{L}(\vec{k} + \vec{p} )^2 \sum_{j} \delta(\vec{k} - \vec{q}_{j}).
\end{equation}
Up to second order in $ \vec{p} $ 
\begin{equation}
	\Delta f_{\rm el} = f_{\rm el} - \left. f_{\rm el} \right|_{\vec{p}=0} = 2 \beta_{\rm A}\phi^2  \sum_{j} [\vec{q}_j \cdot \vec{p} (\vec{q}_j)]^2.
\end{equation}
Let us consider small linear deformations with $ \vec{p}(\vec{k}) = \delta \vec{J} \vec{k} $, with some matrix $ \vec{J} $ with squared eigenvalues $ \lambda_{1}^2 + \lambda_{2}^2 = \Tr (\vec{J}^2) = 1 $ and some small $ \delta $. For a hexagonal lattice, the reciprocal lattice vectors $ \vec{q}_{j} $ form a star with 6-fold symmetry. It can be shown \cite{ref-pfc-mechanical-equilibration} that 
\begin{equation}
\begin{split}
	\Delta f_{\rm el} &= \frac{3}{2}  \delta^2 \beta_{\rm A} \phi^2  
	\left[
	\Tr (\vec{J}^2) + (\Tr\vec{J})^2 + \Tr{(\vec{J}^T \vec{J})} 
	\right] \\
	&> 3 \beta_{\rm A} \phi^2 \delta^2  ( \lambda_{1}^2 + \lambda_{2}^2 ) = 3 \beta_{\rm A} \phi^2 \delta^2 . 
\end{split}
\end{equation}
In order to compare with $ \Delta  f_{\rm ct}^{\text{int}} $ of Eq.~\eqref{eq:interface_excess_elastic_energy}, $ \Delta f_{\rm el} $ needs to be multiplied by the thickness of the interface, which we assume to be two lattice constants $ a = 4\pi/\sqrt{3} $. This gives
\begin{equation}\label{eq:bulk_interface_elastic_energy}
2 a \Delta f_{\rm el}  > 8\sqrt{3} \pi \beta_{\rm A} \phi^2 \delta^2.
\end{equation}
Inserting $ \beta_{\rm A} = 1 $, $ \epsilon_{\rm AB} = 1$ we get an estimate 
\begin{equation}\label{eq:final_ratio}
	\frac{2 a \Delta f_{\rm el}}{\Delta  f_{\rm ct}^{\text{int}} } > 1000,
\end{equation}
which proves that the contribution of the smoothing term to the elastic excitation energies is insignificant.

\end{document}